\documentclass[twocolumn]{aastex631}
\usepackage{epsfig, natbib, graphicx, color}
\usepackage{subfiles}
\usepackage{import}
\usepackage{amsmath}
\usepackage[]{algorithm2e}
\usepackage{longtable}

\graphicspath{{./}{pics/}}

\input epsf 

\usepackage{color}

\def\fom{\Sigma_{\rm FOM}}
\def\fomlim{\Sigma_{\rm FOM, limit}}
\def\sigmakern{\sigma_{\rm kernel}}
\def\sigmasim{\sigma_{\rm sim}}

\shortauthors{Rest et al.}
\shorttitle{SN\,2023ixf and ATClean}

\begin{document}
\title{ATClean: A Novel Method for Detecting Low-Luminosity Transients and Application to Pre-explosion Counterparts from SN\,2023ixf}

\correspondingauthor{S. Rest}
\email{srest2021@gmail.com}
\newcommand{\STSCI}{\affiliation{Space Telescope Science Institute, Baltimore, MD 21218, USA}}

\newcommand{\JHU}{\affiliation{Department of Physics and Astronomy, The Johns Hopkins University, Baltimore, MD 21218, USA}}

\newcommand{\CSJHU}{\affiliation{Department of Computer Science, The Johns Hopkins University, Baltimore, MD 21218, USA}}

\newcommand{\NU}{\affiliation{Center for Interdisciplinary Exploration and Research in Astrophysics (CIERA) and Department of Physics and Astronomy, Northwestern University, Evanston, IL 60208, USA}}

\newcommand{\UC}{\affiliation{School of Physical and Chemical Sciences—Te Kura Matū, University of Canterbury, Private Bag 4800, Christchurch 8140, New Zealand}}

\newcommand{\CIT}{\affiliation{IPAC, California Institute of Technology, 1200 East California Boulevard, Pasadena, CA 91125, USA}}

\newcommand{\CWRU}{\affiliation{Department of Physics, Case Western Reserve University, 2076 Adelbert Rd., Cleveland, OH 44106}}

\newcommand{\UO}{\affiliation{Astrophysics Sub-department, Department of Physics, University of Oxford, Keble Road, Oxford OX1 3RH, UK}}

\newcommand{\QUB}{\affiliation{Astrophysics Research Centre, School of Mathematics and Physics, Queen’s University Belfast, Belfast, BT7 1NN, UK}}

\newcommand{\UHIA}{\affiliation{Institute for Astronomy, University of Hawaii, 2680 Woodlawn Drive, Honolulu, HI 96822, USA}}

\newcommand{\UCC}{\affiliation{Instituto de Astrof\'{\i}sica, Pontificia Universidad Cat\'olica de Chile, Vicu\~na Mackenna 4860, Macul, Santiago, Chile}}

\newcommand{\ESO}{\affiliation{European Southern Observatory, Alonso de C\'ordova 3107, Casilla 19, Santiago, Chile 2}}

\newcommand{\MIA}{\affiliation{Millennium Institute of Astrophysics MAS, Nuncio Monse\~nor S\'otero Sanz 100, Of. 104, Providencia, Santiago, Chile
}}

\newcommand{\UW}{\affiliation{DiRAC Institute and the Department of Astronomy, University of Washington, 3910 15th Avenue NE, Seattle, WA 98195 USA}}

\newcommand{\RO}{\affiliation{Vera C. Rubin Observatory Project Office, 950 N Cherry Ave, Tucson, AZ 95719, USA}}

\newcommand{\ISEF}{ISEF International Fellowship}
\author[0000-0002-3825-0553]{S. Rest}
\CSJHU

\author[0000-0002-4410-5387]{A. Rest}
\JHU\STSCI

\author[0000-0002-5740-7747]{C. D. Kilpatrick}
\NU

\author[0000-0001-5754-4007]{J. E. Jencson}
\CIT

\author[0000-0002-8436-5431]{S. von Coelln}
\CWRU\JHU

\author[0000-0002-7756-4440]{L. Strolger}
\JHU\STSCI

\author[0000-0002-8229-1731]{S. Smartt}
\UO\QUB


\author[0000-0003-0227-3451]{J. P. Anderson}
\ESO\MIA

\author[0000-0003-3068-4258]{A. Clocchiatti}
\UCC\MIA

\author[0000-0003-4263-2228]{D. A. Coulter}
\STSCI

\author[0000-0002-7034-148X]{L. Denneau}
\UHIA

\author[0000-0001-6395-6702]{S. Gomez}
\STSCI

\author[0000-0003-3313-4921]{A. Heinze}
\UW

\author[0000-0003-1724-2885]{R. Ridden-Harper}
\UC

\author[0000-0001-9535-3199]{K. W. Smith}
\QUB

\author[0000-0003-0973-4900]{B. Stalder}
\RO

\author[0000-0003-2858-9657]{J. L. Tonry}
\UHIA

\author[0000-0001-5233-6989]{Q. Wang}
\JHU

\author[0000-0002-0632-8897]{Y. Zenati}
\altaffiliation{\ISEF}
\JHU
\STSCI

\begin{abstract}
In an effort to search for faint sources of emission over arbitrary timescales, we present a novel method for analyzing forced photometry light curves in difference imaging from optical surveys. Our method ``ATLAS Clean'' or ATClean, utilizes the reported fluxes, uncertainties, and fits to the point-spread function from difference images to quantify the statistical significance of individual measurements. We apply this method to control light curves across the image to determine whether any source of flux is present in the data for a range of specific timescales. From ATLAS $o$-band imaging at the site of the Type\, II supernova (SN) 2023ixf in M101 from 2015--2023, we show that this method accurately reproduces the 3$\sigma$ flux limits produced from other, more computationally expensive methods. We derive limits for emission on timescales of 5~days and 80-300~days at the site of SN\,2023ixf, which are 19.8 and 21.3~mag, respectively. The latter limits rule out variability for unextinguished red supergiants (RSG) with initial masses $>$22~$M_{\odot}$, comparable to the most luminous predictions for the SN\,2023ixf progenitor system. We also compare our limits to short timescale outbursts, similar to those expected for Type\, IIn SN progenitor stars or the Type\, II SN 2020tlf, and rule out outburst ejecta masses of $>$0.021~$M_{\odot}$, much lower than the inferred mass of circumstellar matter around SN\,2023ixf in the literature. In the future, these methods can be applied to any forced point-spread function photometry on difference imaging from other surveys, such as Rubin optical imaging. 

\end{abstract}

\section{Introduction}

The final stages of a massive star ($M \gtrsim 8~M_{\odot}$) evolution, especially during the last few years before core-collapse (CC), remain poorly understood. There is clear evidence that some massive stars experience enhanced, eruptive mass-loss episodes at the ends of their lives, forming dense shells of circumstellar material (CSM) in their immediate vicinities \citep[e.g.,][and references therein]{smith2014}. The presence of this material can have dramatic effects on both the appearance of the progenitor star \citep[e.g.,][]{kochanek2012,davies2022}, as well as on the resulting CC supernova (SN). 

As increasing numbers of CCSNe are discovered within days, or even hours, of explosion, a growing body of evidence from multiple observational signatures supports the presence of confined ($<$10$^{14}$~cm) CSM surrounding their progenitor stars, including for the most common class of CCSN, hydrogen-rich Type II SNe (SN~II) arising from red supergiant (RSG) stars \citep[][]{li2011,smartt2009,smartt2015,vandyk2017}. Hydrodynamical modeling of the early light curves of SN\,II \citep[e.g.,][]{morozova2017,morozova2018}, coupled with observations of their rapid rise times \citep[e.g.,][]{andrews2019,dong2021,hosseinzadeh2018,hosseinzadeh2022,pearson2023}, points to early interaction with dense material. In the same vein, many SN\,II exhibit so-called ``flash'' emission features at early times, usually attributed to ionization of or shock-interaction with nearby CSM \citep{gal-yam2014,yaron2017,bruch2021,tartaglia2021}. SN\,II with directly-detected RSG progenitor stars and infrared (IR) constraints on their spectral energy distributions (SEDs) exhibit extremely cool effective temperatures \citep[e.g., SN\,2008bk, 2012A, 2012aw, 2017eaw, and 2023ixf;][]{fraser12,vandyk12,kochanek2012,tomasella13,maund14,kilpatrick2018,jencson2023,kilpatrick2023,soraisam2023,vandyk2023} consistent with a confined shell of dense CSM surrounding the star. Moreover, some progenitor stars of SN\,II that have only optical detections appear to be much less luminous than expected based on their light curve properties \citep[e.g., SN\,2020jfo, 2021yja, 2022acko;][]{Hosseinzadeh22, Kilpatrick2023a, vandyk2023a}, which suggests that the progenitor star may be heavily obscured by extinction from confined CSM.

Several theoretical scenarios to produce enhanced, end-stage mass-loss have been proposed to explain the presence of this material, including instabilities during the final stages of nuclear burning, wave-driven momentum transport, and neutrino-driven mass loss \citep[e.g.,][]{heger1997,yoon2010,arnett2011,quataert2012,shiode2013,moriya2014,shiode2014,smith2014,woosley2015,fuller2017,wu2021}. Precursor emission appears to be common leading up to the strongly interacting SN\,IIn (those with persistent ``narrow'' emission lines indicative of very massive CSM shells; e.g., \citealp{ofek2014,strotjohann2021}). They have only been directly observed for more normal SN~II progenitors in a few instances \citep{kilpatrick2018,jacobson-galan2022}, while some SN~II progenitors appear to be quiescent in their final years \citep{johnson2018}. Recent modeling efforts suggest that SN precursor emission can be explained as eruptive outbursts with $\sim$0.1--1~$M_{\odot}$ of material ejected at $\sim$100--1000~km~s$^{-1}$, or for lower luminosity precursors ($\lesssim$10$^{41}$~erg~s$^{-1}$), steady enhanced winds at high mass-loss rates up to $\sim$1~$M_{\odot}$~yr$^{-1}$ \citep{matsumoto2022}.

Future efforts to constrain these scenarios using optical time-domain surveys such as the Young Supernova Experiment \citep{Jones2021}, the Zwicky Transient Facility \citep{Bellm19}, and the Vera C. Rubin Observatory's Legacy Survey of Space and Time \citep{Ivezic19} will require algorithms tailored to detecting the low-luminosity counterparts expected for these mechanisms on varying timescales. The vast majority of existing studies of pre-SN massive star evolution are performed in retrospect, but a flexible algorithm that can efficiently characterize sources of transient emission and look for short timescale outbursts or long-period massive star variability similar to the SN\,2023ixf progenitor \citep{jencson2023,kilpatrick2023,soraisam2023,vandyk2023,Ransome2024} could provide a ``SN warning system'' for targeted follow up {\it before} these events take place.

Here, we present the results of a search for optical precursor emission or pre-SN outbursts for the Type II SN\,2023ixf using extensive coverage by the ATLAS survey \citep{tonry2018}. The SN was discovered in M101 ($D = 6.85\pm0.15$~Mpc; $\mu = 29.18$~mag; \citealp{riess2022}) on UT 2023 May 19.72 within hours of the explosion by \citet{Itagaki2023}. Intensive monitoring of SN\,2023ixf already shows evidence of dense, close-in CSM \citep{bostroem23,hosseinzadeh2023,hiramatsu2023,jacobson-galan2023,smith2023}. As one of the nearest SN~II in recent decades, this event offers a golden opportunity to place stringent constraints on precursor activity for an SN that will be exceptionally well-studied throughout its evolution. For this analysis, we use a novel method called ``ATLAS Clean'' or ATClean for analyzing forced photometry in difference imaging obtained from the ATLAS survey to conduct a search for faint transients over arbitrary timescales.

\section{Observations}\label{sec:observations}

The ATLAS survey is a network of four identical, small-aperture telescope units in both hemispheres. Two telescopes in Hawaii are located on Haleakala and Mauna Loa, which have been operating for more than seven years, while the telescopes in El Sauce (Chile) and Sutherland (South Africa) were commissioned more recently in 2022. Since SN\,2023ixf has a Declination (Decl.) of $\delta= 54.31165528^{\circ}$ \citep[J2000;][]{Itagaki2023}, all light curve data we use for this analysis is from the northern telescopes in Hawaii. ATLAS observes in the orange ($o$) and cyan ($c$) filters, similar to the Pan-STARRS $r+i$ and $g+r$ filters, respectively. As described in \cite{tonry2018}, the ATLAS units are Schmidt telescopes with a 0.5m clear aperture, and use a 10.5K $\times$ 10.5K STA-1600 detectors with a plate scale of $\sim$1.86\arcsec~pixel$^{-1}$. ATLAS's resulting field-of-view is approximately 28.9~deg$^2$. Typically, ATLAS has a cadence of 2 days, with 4 exposures at each epoch separated by 10 minutes to search for moving objects such as asteroids.

We obtain the ATLAS light curve for SN\,2023ixf using the ATLAS Forced Photometry server \citep{shingles2021}. 
Besides the SN\,2023ixf light curve, we also obtain a set of control light curves, which are located in a circular pattern around the location of SN~2023ixf (right ascension (R.A.) $210^\text{h}54^\text{m}38^\text{s}.42$ and Decl. $54^{\circ}18'41.96\arcsec$), 
with a radius of 34\arcsec. They are located sufficiently far away from SN\,2023ixf to be deemed independent---that is, they contain no real astrophysical flux---yet close enough such that they are impacted by similar systematic noise sources. We designate these light curves as \textit{controls} for further identification and removal of bad measurements without using any part of the SN light curve itself. The positions of each control light curve as well as their offsets from the SN position are listed in Table~\ref{table:controls} and illustrated in Fig.~\ref{fig:cutouts}. 

We proceed with our analysis by taking into account only the $o$-band ATLAS photometry. Due to the lower cadence of the $c$-band \citep{tonry2018}, the constraints derived from the photometry are inherently less stringent, so we focus on $o$-band measurements only.

\section{Data Reduction}

We utilize ATLAS forced photometry light curves obtained from 2015---2023 difference imaging to search for low-luminosity, transient emission signatures at the site of SN\,2023ixf in M101. All difference imaging and photometry are obtained as described in \citet{tonry2018}. Here we describe our procedure to validate and ``clean" the forced photometry light curves described above, in preparation for constraining the presence of faint, arbitrary timescale emission within them. 
Python scripts used to download, clean, bin, and otherwise modify or analyze the ATLAS forced photometry light curves used in this paper are available at the \dataset[ATClean repository]{https://github.com/srest2021/atclean} on GitHub, as well as through the version 4 \dataset[Zenodo release]{https://zenodo.org/records/10914544}. 

\subsection{Cleaning Individual Measurements}
\label{cleaning1}

Systematic residuals in the light curve caused by instrument artifacts, close bright objects, or reduction artifacts can mimic pre-SN eruptions from SNe, and it is therefore necessary to distinguish between evidence for actual transient flux and false positives arising from other sources. To ensure the reliability of ATLAS light curves for analysis and interpretation, we validate the accuracy of individual detections through a series of cuts and corrections involving the reported fluxes of the SN and its control light curves, as well as statistical measures such as the nominal uncertainties and the chi-squares of the point-spread function (PSF) fit. In the following sections, we detail our procedure for identifying and removing systematic residuals, correcting the data, and applying these cuts in a systematic way to clean the photometry and obtain photometric accuracy. 

Our approach begins with the acquisition of a set of sixteen forced photometry control light curves around SN~2023ixf, as described in Section~\ref{sec:observations}.
Crucially, the following detailed cuts and corrections are applied not only to the entire ATLAS $o$-band light curve of SN~2023ixf, but also to each of its sixteen control $o$-band light curves.

\begin{table*}[htb!]
    \centering
    \caption{SN\,2023ixf and Control Light Curve Locations}
    \begin{tabular}{lccccc} 
\hline
Control ID &  R.A. & Decl. & Radius & R.A. Offset & Decl. Offset \\
& \textbf{(deg)} & \textbf{(deg)} & \textbf{(\arcsec)} & \textbf{(deg)} & \textbf{(deg)} \\
    \hline\hline
    1 & 210.92562821 & 54.31526951 & 34.0 & 0.01495696 & 0.00361423 \\
    2 & 210.92211881 & 54.31833351 & 34.0 & 0.01144756 & 0.00667823 \\ 
    3 & 210.91686662 & 54.32038081 & 34.0 & 0.00619537 & 0.00872553 \\ 
    4 & 210.91067125 & 54.32109972 & 34.0 & 0.00000000 & 0.00944444 \\ 
    5 & 210.90447588 & 54.32038081 & 34.0 & -0.00619537 & 0.00872553 \\ 
    6 & 210.89922369 & 54.31833351 & 34.0 & -0.01144756 & 0.00667823 \\ 
    7 & 210.89571429 & 54.31526951 & 34.0 & -0.01495696 & 0.00361423 \\ 
    8 & 210.89448196 & 54.31165528 & 34.0 & -0.01618929 & 0.00000000 \\ 
    9 & 210.89571429 & 54.30804105 & 34.0 & -0.01495696 & -0.00361423 \\
    10 & 210.89922369 & 54.30497705 & 34.0 & -0.01144756 & -0.00667823 \\ 
    11 & 210.90447588 & 54.30292975 & 34.0 & -0.00619537 & -0.00872553 \\
    12 & 210.91067125 & 54.30221083 & 34.0 & 0.00000000 & -0.00944444 \\
    13 & 210.91686662 & 54.30292975 & 34.0 & 0.00619537 & -0.00872553 \\ 
    14 & 210.92211881 & 54.30497705 & 34.0 & 0.01144756 & -0.00667823 \\ 
    15 & 210.92562821 & 54.30804105 & 34.0 & 0.01495696 & -0.00361423 \\
    16 & 210.92686054 & 54.31165528 & 34.0 & 0.01618929 & 0.00000000 \\ 
    \hline
    \end{tabular}
    \label{table:controls}
\end{table*}

\begin{figure}[htb!]
    \centering
\includegraphics[width=0.37\textwidth]{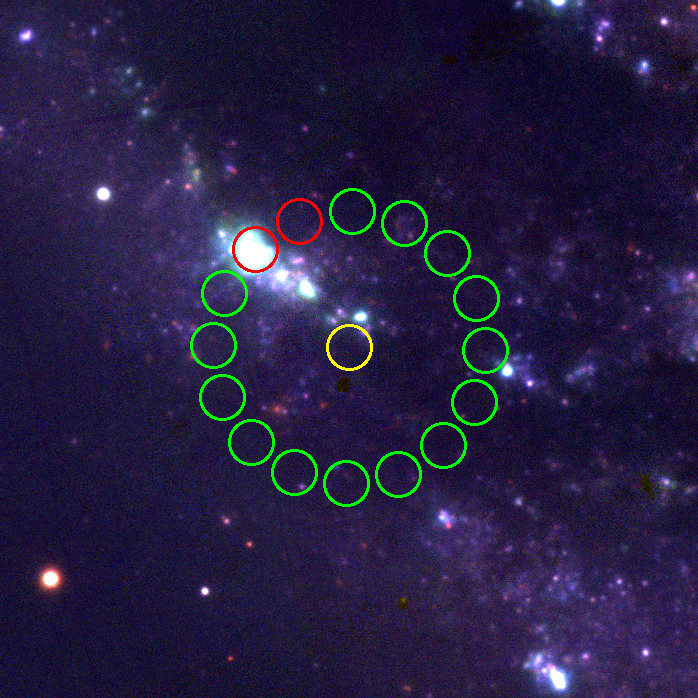}
\caption{Position of SN~2023ixf (yellow circle) and the control light curves (red and green circles). The color image is created using Pan-STARRS1 {\it gri} images. The control light curves are numerated counter-clockwise, with \#4 at the top. Control light curve \#2 (leftmost red circle) is on top of a bright emission line region, and shows significantly more scatter than any of the other light curves (see Appendix Fig.~\ref{fig:all_controls}), and we therefore exclude it from our sample of control light curves since it is not representative of the pre-SN light curve of SN~2023ixf. Control light curve \#3 (rightmost red circle) also showed an offset in the early seasons, most likely due to some issue with the template image at that position, and for simplicity we also remove this light curve from our control light curve sample.}
    \label{fig:cutouts}
\end{figure}

\subsubsection{Uncertainty Cut}

We start with the light curves at the site of SN\,2023ixf and its controls in $o$-band, including flux measurements $f$ in $\mu$Jy and their nominal uncertainties $\sigma_{f,0}$ in $\delta \mu$Jy. For every data point (i.e., every individual epoch of imaging), we either cut or keep that measurement according to its corresponding uncertainty value. 

To determine the right uncertainty cutoff value for any egregious outliers, we calculate the typical uncertainty of bright stars just below the saturation limit as 160~$\mu$Jy. (Note that this value is set as a constant in our pipeline.) Then, for every data point in the SN light curve and control light curves, any measurements in the light curve with $\sigma_{f,0} > 160$ $\mu$Jy are cut. Approximately 5.26\% of SN\,2023ixf's $o$-band measurements were removed by this cut.

\subsubsection{Uncertainty Correction}

The nominal uncertainties $\sigma_{f,0}$ sometimes underestimate the true uncertainties, which may be due to unaccounted-for covariances arising from convolution in the deprojection and difference image steps, or due to other systematic sources of noise. We therefore estimate the ``true'' convolution-corrected uncertainties $\sigma_f$ as the sum of the squares of the nominal uncertainties $\sigma_{f,0}$ and the systematic uncertainties $\sigma_{f,{\rm sys}}$:

\begin{equation}
    \label{equation:uncertest}
    \sigma_f^2 = \sigma_{f,0}^2 + \sigma_{f,{\rm sys}}^2
\end{equation}

We first estimate the systematic uncertainties $\sigma_{f,{\rm sys}}$ with respect to the control light curves, under the assumption that they contain no real astrophysical flux. Temporarily and for the sole purpose of this estimation, we disregard the previous uncertainty cut of 160~$\mu$Jy and instead apply a cut of 20 to the reduced chi-square of the PSF fits to our source apertures in the difference imaging, i.e., $\chi^2_\text{PSF}$. Then, the typical uncertainty $\sigma_{f,{\rm typical}}$ is estimated as the median of the remaining nominal uncertainties $\sigma_{f,0}$ with $\chi_{\text{PSF}}^2\leq20$ across all control light curves. We also calculate the empirical 3$\sigma$-clipped standard deviation  $\sigma_{f,{\rm empirical}}$ of these remaining measurements. If $\sigma_{f,{\rm empirical}} \le \sigma_{f,{\rm typical}}$, $\sigma_{f,{\rm sys}}$ is set to 0; otherwise, $\sigma_{f,{\rm sys}}$ is estimated by the following equation:

\begin{equation}
    \sigma_{f,{\rm empirical}}^2 = \sigma_{f,{\rm typical}}^2 + \sigma_{f,{\rm sys}}^2
\end{equation}

For SN\,2023ixf, we estimate $\sigma_{f,{\rm sys}} = 10.89$ $\delta\mu$Jy. This extra systematic noise $\sigma_{f,{\rm sys}}$ is then added to the SN and control light curves' nominal uncertainties $\sigma_{f,0}$ in quadrature to obtain $\sigma_f$ per Equation \ref{equation:uncertest}.


We ultimately increase the typical uncertainty $\sigma_{f,{\rm typical}}$ by 15.27\% from $19.00$ $\delta\mu$Jy to $21.90$ $\delta\mu$Jy. The true uncertainties $\sigma_f$ are then used in replacement of $\sigma_{f,0}$ throughout the rest of the following cuts.

\subsubsection{PSF Chi-Square Cut}
Each data point in the SN and control light curves is now either cut or kept according to the corresponding reduced chi-square of the difference image point-spread function (PSF) fit, $\chi^2_\text{PSF}$. In order to choose the best cutoff point, we evaluate the effectiveness of a range of possible $\chi^2_\text{PSF}$ cuts using the measures \textit{loss} and \textit{contamination}, then choose the cut which optimally minimizes both measures.

Before calculating these values, a deciding factor independent of $\chi^2_\text{PSF}$ must be chosen and used to classify each measurement as either ``good" or ``bad.'' We define a ``good" measurement to have $\frac{f}{\sigma_{f}} \le 3$; otherwise, the measurement is classified as ``bad.'' The loss $\mathcal{L}$ of any arbitrary $\chi^2_\text{PSF}$ cut can therefore be defined according to the number of cut and/or good measurements within the the control light curves:
\begin{equation}
    \mathcal{L} = \frac{N_{\text{good, cut}}}{N_{\text{good}}}
\end{equation}
Similarly, we define the contamination $\mathcal{C}$ of the cut according to the number of kept and/or bad measurements within the control light curves:
\begin{equation}
    \mathcal{C} = \frac{N_{\text{bad, kept}}}{N_{\text{kept}}}
\end{equation}

\begin{figure}[htb!]
        \includegraphics[scale=1, width=0.49\textwidth]{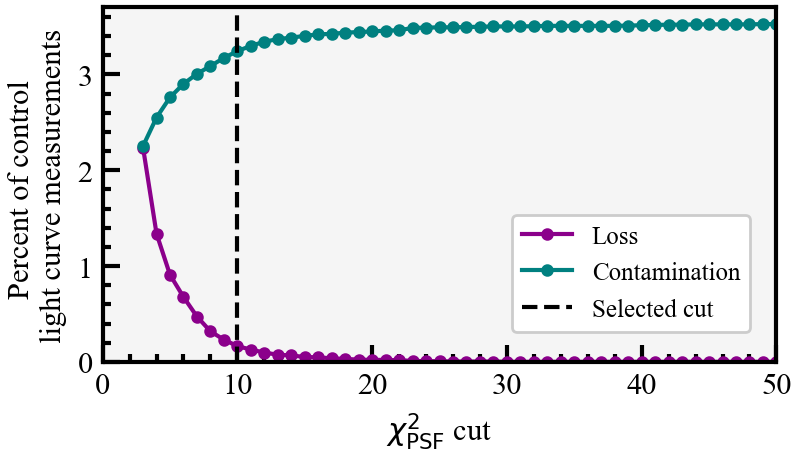}
        \caption{
            Loss $\mathcal{L}$ (purple) and contamination $\mathcal{C}$ (teal) within the control light curves for a range of chi-square cuts from $\chi^2_\text{PSF}$ = 3 to 50. In the dashed black line, the chosen cut at $\chi^2_\text{PSF}$ = 10.
        }
        \label{fig:chisquare_cut}
        \centering
\end{figure}

In this way, we can calculate loss and contamination for a series of cuts from $\chi^2_\text{PSF}$ = 3 to 50 and choose the cut which best minimizes both loss of good measurements and contamination of bad measurements within the control light curves. For SN~2023ixf, we prioritize the minimization of loss. Figure~\ref{fig:chisquare_cut} displays $\mathcal{L}$ and $\mathcal{C}$ with respect to the percent of control light curve measurements for this range of cuts. Evidently, $\mathcal{L}$ begins to increase rapidly at $\chi^2_\text{PSF}$ = 10 as it approaches 0. We therefore select $\chi^2_\text{PSF}$ = 10 as the optimal cut for SN\,2023ixf. For $\chi^2_\text{PSF}$ = 10, $\mathcal{C} = 3.24\%$ and $\mathcal{L} = 0.16\%$, thus effectively minimizing the loss of good measurements.

Any measurements in the light curve with $\chi^2_\text{PSF} > 10$ are cut. Approximately 2.48\% of SN\,2023ixf's $o$-band measurements were removed by this cut.

\subsubsection{Control Light Curve Cut}

We proceed with the next cut by again operating under the assumption that the control light curves contain no real astrophysical flux. For a given SN epoch, we can calculate the 3$\sigma$-clipped average of the corresponding sixteen control flux measurements falling within the same epoch. Given the expectation of control flux consistent with zero, the statistical properties accompanying the 3$\sigma$-clipped average enable us to identify problematic epochs, which we subsequently cut from both the SN and control light curves. 

We cut any epochs in the light curves for which the 3$\sigma$-clipped average's statistics fulfill any of the following criteria: (a) $\chi^2 > 2.5$, (b) $|\frac{f}{\sigma_{f}}| > 3$, (c) the number of measurements clipped in this epoch is greater than 2, and (d) the number of measurements averaged in this epoch is less than 4. Approximately 3.72\% of SN\,2023ixf's $o$-band measurements were cut by this procedure. 

\begin{figure}[htb!]
        \includegraphics[scale=1, width=0.49\textwidth]{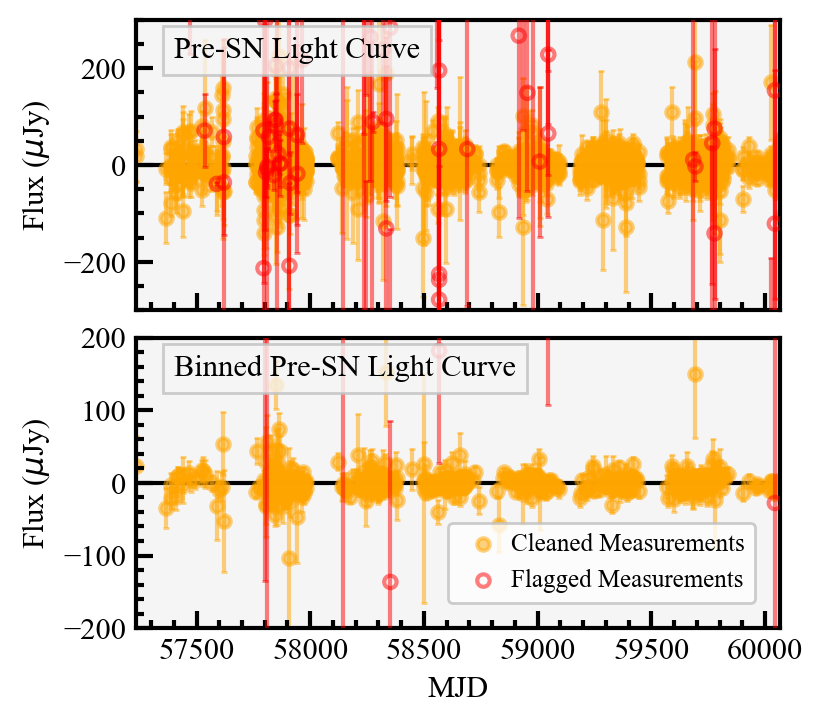}
        \caption{
            \textbf{Top}: The pre-SN ATLAS $o$-band light curve, with flagged measurements in open red circles and kept measurements in closed orange circles. 8.26\% of all individual measurements were flagged. \textbf{Bottom}: The pre-SN binned ATLAS $o$-band light curve, with flagged measurements in open red circles and kept measurements in closed orange circles. 6.06\% of all binned days were flagged.
        }
        \label{fig:cleaned_avg}
        \centering
\end{figure}

In total, approximately 8.26\% of the measurements in SN\,2023ixf's $o$-band ATLAS light curve were removed by the previous cuts (uncertainty, PSF chi-square, and control light curve cut). The top panel of Fig. \ref{fig:cleaned_avg} displays the pre-SN ATLAS $o$-band light curve, as well as the removed measurements.

\subsection{Correcting for ATLAS Template Changes}

The ATLAS survey employs an image subtraction pipeline using ``all-sky reference images'' or ``wallpapers'' to generate difference imaging and the corresponding difference flux light curves \citep{tonry2018}. As the difference image reference templates are periodically replaced, Poisson noise associated with those templates may be added to the flux to varying degrees, and step discontinuities may appear in a SN light curve when these replacements occur. For the SN\,2023ixf field, these transitions happened at MJD 58417 from the wallpaper version 1 to 2, and at MJD 58882 from the wallpaper version 2 to 3 \citep{tonry2018}. We correct for these step discontinuities in the SN\,2023ixf light curve by calculating the corrective flux necessary to offset the discontinuities present in the regions surrounding the template transition dates. We add 5 $\mu$Jy to flux with MJD$>$58882 and subtract 5 $\mu$Jy from all other epochs.


\subsection{Cleaning Binned Measurements}



We now segment the cleaned SN and control light curves into time bins of 1 day each, with the aim of further identifying bad epochs and refining our dataset for enhanced accuracy. We establish the start of the first bin as the onset of the first day with recorded measurements, and each bin is exactly 24 hours long. Since the ATLAS survey takes approximately 4 exposures within 2 hours of each other every 2 days, we average together approximately 4 measurements per bin. Note that out of these 4 exposures, only measurements not removed in the previous cuts are passed to the 3$\sigma$-clipped average. 

We cut any bins for which the 3$\sigma$-clipped average’s statistics fulfill any of the following criteria: (a) the number of measurements not flagged in the previous cuts is less than or equal to 2, (b) the number of measurements clipped in this epoch is greater than 1, or (c) the number of measurements averaged in this epoch is less than 2. 

The bottom panel of Fig.~\ref{fig:cleaned_avg} displays the final cleaned and binned pre-SN ATLAS $o$-band light curve, as well as the removed measurements. Approximately 6.06\% of SN\,2023ixf's $o$-band binned days were cut by this procedure.

\section{Analysis}

\begin{figure}[htb!]
        \includegraphics[scale=1, width=0.49\textwidth]{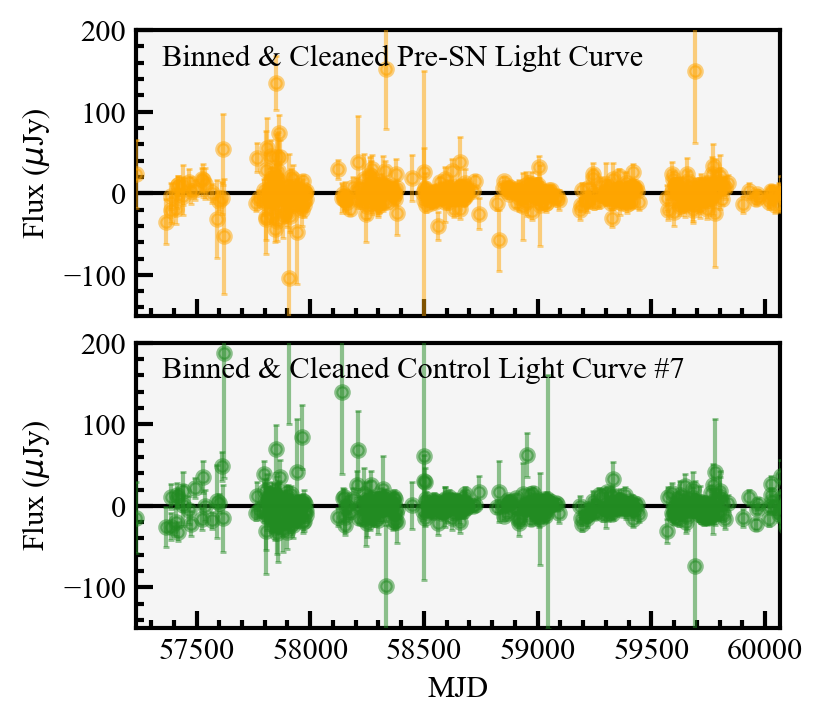}
        \caption{
            \textbf{Top}: In orange, the pre-SN ATLAS $o$-band light curve, cleaned and binned. \textbf{Bottom}: In dark green, the ATLAS $o$-band light curve of the example control light curve \#7, cleaned and binned.
        }
        \label{fig:avg_control}
        \centering
\end{figure}

Pre-SN outbursts have been previously observed in SN\,II-P, for example 130~days prior to 2020tlf \citep{jacobson-galan2022}, motivating systematic studies for pre-SN activity when these events are sufficiently close and time-domain surveys are sufficiently deep to place meaningful constraints these outbursts. Therefore, we proceed to search for outbursts on similar timescales before SN\,2023ixf with similar characteristic durations in the outburst. ATLAS monitored the field of SN\,2023ixf in the $c$ and $o$ bands for approximately seven years prior to explosion. In the top panel of Fig.~\ref{fig:avg_control}, we display the pre-explosion binned and cleaned ATLAS forced photometry $o$-band light curve for SN\,2023ixf. We also highlight a single control light curve taken 34\arcsec\ away from the location of the SN for illustration. The binned and cleaned light curve of the selected control location is displayed in the bottom panel of Fig.~\ref{fig:avg_control}.

\subsection{Contamination \& Efficiency Determination}
\label{contameffmjdranges}

When determining the presence of a pre-SN outburst in an ATLAS light curve, it is important to account for contamination caused by factors such as instrument and reduction artifacts or nearby bright objects, while simultaneously remaining sensitive to faint eruptions. The challenge, however, is to do so in a well-defined, robust, and quantitative way. With our series of control light curves (previously described in Section~\ref{cleaning1}), we can perform a quantitative efficiency and contamination analysis. We again assume that the control light curves contain no real astrophysical flux (barring any unlikely coincidental unrelated transient or variable source), and should therefore be consistent with zero within the uncertainties. For a given pre-SN outburst detection algorithm, we thus utilize the control light curves in two ways.

The first way is to calculate \textit{contamination}. Any false positive by the detection algorithm on an unchanged control light curve may be classified as contamination. We can characterize contamination by either (a) the total number of false positives across all control light curves or (b) the number of control light curves that contain any false positives. 

The second way is to calculate \textit{efficiency}. Our control light curves contain similar instrument and reduction artifacts to the SN light curve, as well as non-astrophysical false positives such as cosmic ray hits and satellite tracks with similar frequency and instrumental statistics to the SN light curve. This means that if we add a simulated outburst event to one of these control light curves, not only have we simulated the event itself, but the resulting simulated light curve is also reflective of all the systematic trends present in the pre-SN light curve itself. We can therefore run our detection algorithm on control light curves containing simulated events and determine which events were successfully detected within the given parameter space (i.e., the simulated event's timescale, the parameters of the detection algorithm, etc., to be discussed further in subsequent sections). 

Adding simulated events to the control light curves offers a notable advantage; while the light curves closely emulate real-world scenarios, the efficiency and contamination measures remain entirely independent from the SN light curve. In the context of validating our algorithm, we therefore carefully examine each control light curve to confirm consistency with the pre-SN light curve and each other. 

We identify consistently nonzero flux and/or poor quality in control light curves \#2 and \#3 (illustrated alongside the rest of the control light curves in Appendix Fig.~\ref{fig:all_controls}). As depicted in Fig.~\ref{fig:cutouts}, the position of control light curve \#2 is located on top of a bright star-forming region of the galaxy, a property inconsistent with the level and structure of background flux closer to the site of SN\,2023ixf. Similarly, while control light curve \#3 exhibits significantly better quality than control light curve \#2, it also exhibits excess flux throughout the first three observations. Control locations \#2 and \#3 thus warrant exclusion from our analysis as they negate our critical assumption that no real astrophysical flux is present in their light curves.

We further observe several observation seasons (i.e., year-long MJD ranges) within the SN and control light curves with inconsistent properties that could negatively impact efficiency. If bright simulated events within certain MJD ranges fail to reach maximum efficiency at any magnitude, we are unable to proceed with accurate further analysis (such as measuring the magnitude at which an efficiency curve crosses 80\%), and these ranges may warrant exclusion. We therefore identify two criteria for exclusion of a certain MJD range from the main analysis: (a) the specified MJD range is designated as having significantly different properties (e.g., average noise) than other seasons, or (b) bright simulated events within the MJD range do not reach 100\% efficiency. 

As demonstrated through examination of the cleaned and binned pre-SN and example control light curves in Fig.~\ref{fig:avg_control}, the quality of \textit{both} light curves exhibits a marked increase after the initial two observation seasons (MJD$<$58120), and the first season (MJD$\leq$57622) exhibits significantly poorer cadence than the following seasons. Further inspection of the rest of the control light curves (see Appendix Fig.~\ref{fig:all_controls} for illustration) reveals similar quality degradation within the problem seasons, which may be attributed to poorer telescope and pipeline performance. These seasons constitute valid candidates for exclusion by criterion (a).

We observe an increase in non-detections of simulated events within the empty gaps between observation seasons, clearly due to the lack of measurements. We also observe edge effects with significantly more non-detections in MJD ranges very close to certain observation gaps, as well as within the range 59886$\leq$MJD$\leq$59970. By criterion (b) these ranges may warrant exclusion from the main analysis.

\begin{figure}[tb!]
    \centering
\includegraphics[width=0.49\textwidth]{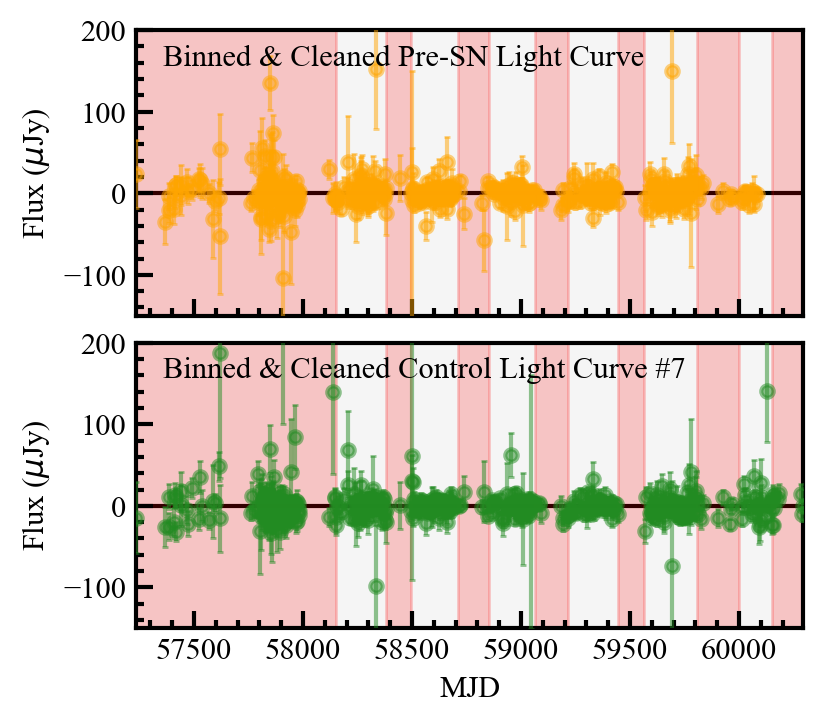}
    \caption{In red, the MJD ranges excluded from our main analysis: MJD$<$58120, 59886$\leq$MJD$\leq$59970, and the fine-tuned observation gaps.}
    \label{fig:mjd_ranges}
\end{figure}

We proceed by excluding the first two observation seasons, the observation gaps with fine-tuned additional buffers on either side where needed, and the aforementioned range 59886$\leq$MJD$\leq$59970 from our main search for pre-SN outbursts in the SN\,2023ixf light curve, and highlight them red in Fig.~\ref{fig:mjd_ranges}. Out of a total of 1938 individual measurements in the ATLAS $o$-band light curve, we include 1306 (67.39\%); in our binned $o$-band light curve, we include 318 of 462 non-empty epochs (68.83\%). We explain these exclusions in more detail in Appendix Section~\ref{s2}. 

\subsection{Pre-SN Outburst Detection Algorithm}

Our primary objective is to identify the presence (or lack thereof) of pre-SN outbursts in the SN\,2023ixf light curve with some degree of certainty. The subsequent analysis and characterization of corresponding ``bumps'' of emission extends beyond the scope of the detection algorithm and constitutes a follow-up analysis.  We therefore base our analysis on the most fundamental property of these outbursts, that is, that they are above zero for a certain continuous timescale. An effective approach to amplifying a signal occurring at a certain timescale is to convolve it with a rolling Gaussian of a similar timescale. Throughout this section, we describe that procedure and its application to the analysis of SN\,2023ixf with the goal of constraining the presence of outbursts in the pre-SN light curve.

In the case of pre-SN outbursts, the signal under consideration is not the flux itself but rather the extent to which the flux surpasses zero with respect to its uncertainty. We therefore define a figure of merit (FOM; hereafter referred to as $\fom$) as the flux-to-uncertainty ratio convolved with a weighted rolling Gaussian sum (hereafter referred to as a ``rolling Gaussian"). We divide this rolling Gaussian by a binary function $\mathcal{B}$, that is, 1 if the epoch contains a flux measurement, and 0 if it does not. Thus we normalize the $\fom$ with respect to the number of significant measurements present. 
\begin{equation}
    \fom = \frac{\int \mathcal{N}(\sigmakern) * f/\sigma_{f} }{\int \mathcal{N}(\sigmakern) * \mathcal{B}}
\end{equation}
\begin{equation}
    \mathcal{B} = 
    \begin{cases}
        1 & \text{if measurement in epoch}\\
        0 & \text{if no measurement in epoch}
    \end{cases}
\end{equation}
Crucially, the rolling Gaussian should be characterized by a standard deviation $\sigmakern$ similar to the timescale of the target pre-SN outburst, in order to best amplify the signal. 

\subsection{Detection Limits}
\label{deteclimits}

To determine whether a given signal is real, we first establish a detection limit $\fomlim$ for a given $\sigmakern$. This detection limit is essential not only to evaluation of whether or not an outburst is present in the pre-SN light curve, but also to measurement of algorithmic effectiveness via efficiency and contamination when simulating these events. As such, we aim to carefully devise a method that determines detection limits most representative of each parameter space. With this goal in mind, we convolve the control light curves---which provide an opportunity to set a threshold for any real outbursts \textit{independent} of the pre-SN light curve---with rolling Gaussians of varying kernel sizes $\sigmakern$. Next, we calculate possible detection limits for each $\sigmakern$, and evaluate their validity by computing the resulting contamination. In this context, we define contamination to be the number of control light curves with any false positives, where a single false positive is defined by a continuous sequence of $\fom$ above $\fomlim$.

\begin{figure}[tb!]
\includegraphics[width=0.49\textwidth]{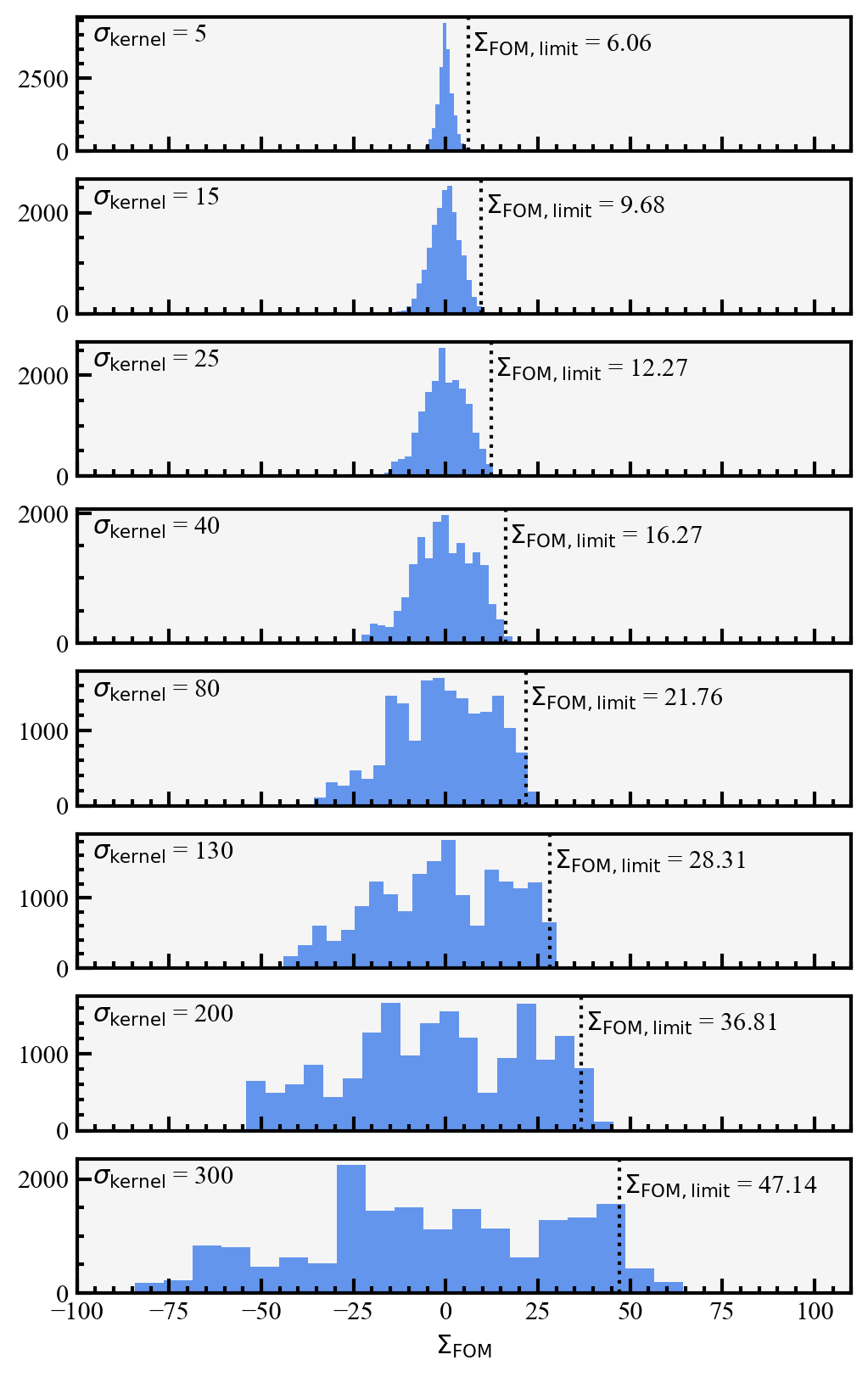}
    \caption{For each $\sigmakern$, the control light curve $\fom$ distribution. In dotted black lines, the corresponding best $\fomlim$ for the selected $\sigmakern$.} 
    \label{fig:fom_hists}
\end{figure}

Our first approach to the problem was to take the control light curve $\fom$ distribution for each $\sigmakern$, apply a $3\sigma$-clipped average, and then scale the obtained standard deviation to get a preliminary detection limit. However, this method inherently assumes $\fom$ distributions that are consistently Gaussian-shaped.  Fig.~\ref{fig:fom_hists} illustrates the discussed $\fom$ distribution for each $\sigmakern$; evidently, while the smaller kernel sizes up to $\sigmakern$ = 40 days produce Gaussian-shaped distributions centered around $\fom$=0, symmetry and Gaussian shape diminishes as the kernel size increases. Therefore, the scaled standard deviation becomes less meaningful for larger kernel sizes. This problem becomes further apparent during the calculation of contamination (see Appendix Section~\ref{deteclimitalt}), such that the smaller kernel sizes exhibit greater contamination than the larger ones.

Ideally, contamination should be uniform across each $\sigmakern$. Therefore, we must account for contamination when calculating the best possible detection limit for a certain $\sigmakern$. We select a target contamination value of two positive control light curves for each $\sigmakern$ and its corresponding $\fomlim$. 

For each $\sigmakern$, we use the \textit{bisection method} with $N_{\text{max}} = 15$ steps to find $\fomlim$ which best fulfills the aforementioned criterion. As a range of possible detection limits, we set the lower endpoint $\beta$ of the range to 0 $\fom$ and the upper endpoint $\alpha$ to the maximum $\fom$ value across all control light curves, resulting in an accuracy of $\frac{\alpha-\beta}{2^{N_{\text{max}}}}$. The pseudocode for the algorithm may be found in Appendix Section~\ref{bisectionmethod}. 

\begin{figure}[htb!]
\includegraphics[width=0.49\textwidth]{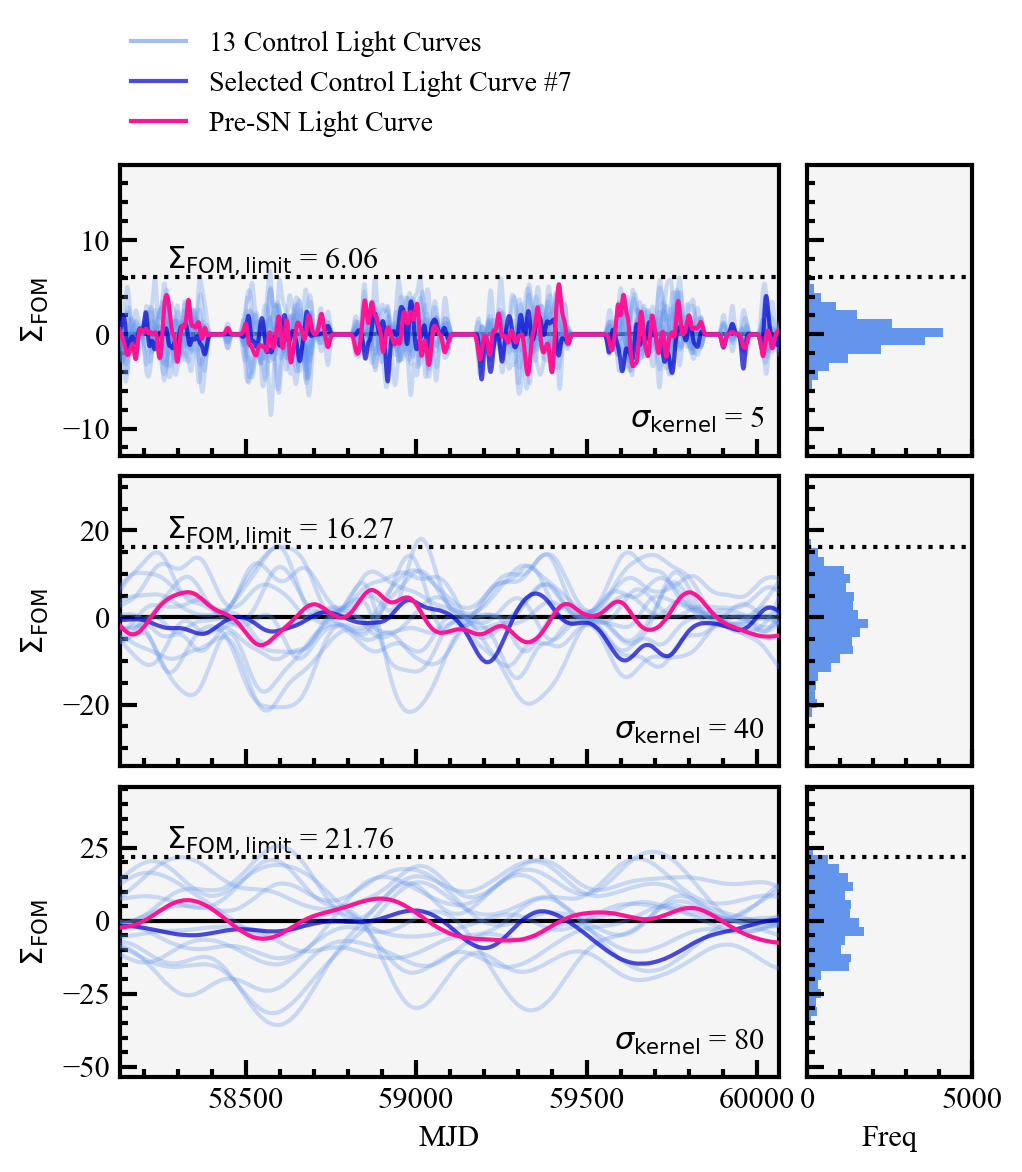}
    \caption{From top to bottom, the $\fom$ of the light curves with $\sigmakern$ = 5, 40, and 80 days, and the corresponding $\fomlim$ in the dotted black lines. To the right, the control light curve $\fom$ distribution for its corresponding $\sigmakern$. The pre-SN $\fom$ (pink) stays below the $\fomlim$ for each $\sigmakern$.}
    \label{fig:all_foms2}
\end{figure}

We depict the best corresponding $\fomlim$ for each $\sigmakern$ in Fig.~\ref{fig:fom_hists}. Figure~\ref{fig:all_foms2} further displays, for select $\sigmakern$, the $\fom$ of the example control light curve \#7, the remaining kept control light curves, and the pre-SN light curve (see Appendix Fig.~\ref{fig:all_foms} for the $\fom$ of all analyzed $\sigmakern$). As depicted, the pre-SN light curve $\fom$ remains below  the corresponding $\fomlim$ for each $\sigmakern$.

\subsection{Simulating Pre-SN Eruptions Using Gaussians}

We inject simulated outbursts into the control light curves to test our detection algorithm and calculate efficiency. Given a rolling Gaussian with a certain $\sigmakern$, we model the target pre-SN outbursts by injecting Gaussian bumps over similar timescales and with varying peak apparent magnitudes. Then, we convolve the resulting simulated flux with the rolling Gaussian and observe if the simulated $\fom$ rises above the detection limit $\fomlim$. 

\begin{figure}[bht!]
    \includegraphics[width=0.49\textwidth]{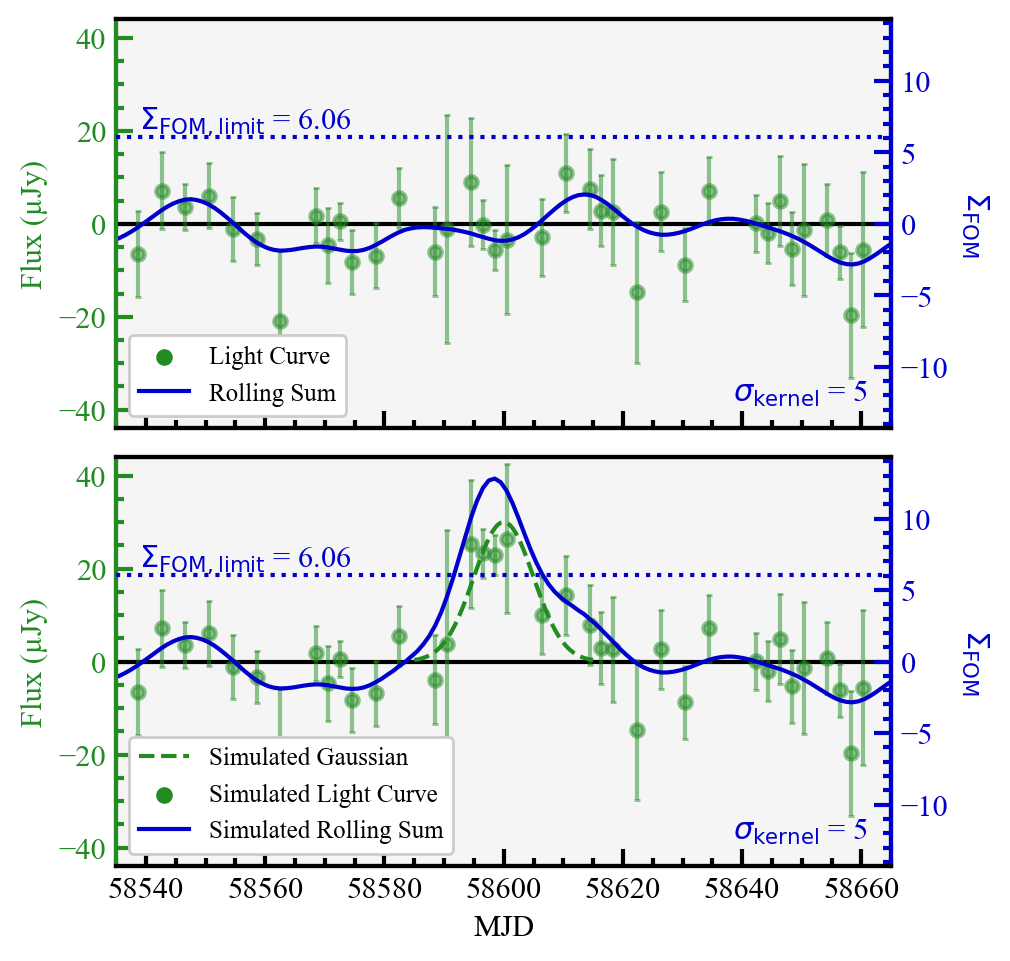}
    \caption{
        \textbf{Top}: In the green points, the example control light curve \#7 flux. The corresponding $\fom$ for $\sigmakern$ = 5 days in the solid blue line, and its $\fomlim$ = 6.06 in the dotted blue line. The $\fom$ does not cross $\fomlim$ at any point. \textbf{Bottom}: In the dashed green line, the shape of the simulated Gaussian with $\sigmasim$ = 5 days peaking at $m_\text{peak}$ = 20.2 mag and 58600 MJD. In the green points, the resulting flux after injection of the simulated Gaussian. The corresponding simulated $\fom$ for $\sigmakern$ = 5 days in the solid blue line, and its $\fomlim$ = 6.06 in the dotted blue line. The $\fom$ crosses $\fomlim$ at approximately 58592 MJD.
    }
    \label{fig:simbump_5}
\end{figure}

In Fig.~\ref{fig:simbump_5}, we inject an example simulated Gaussian with kernel size $\sigmasim$ = 5 days peaking at $m_\text{peak}$ = 20.2 mag. Then, we convolve it with a rolling Gaussian of kernel size $\sigmakern$ = 5 days. We observe that the control light curve $\fom$ successfully crosses the detection limit $\fomlim$ after injection of the simulated event. 


\subsection{Simulating Pre-SN Eruptions Using a Model}

We inject short timescale emission into our light curves to simulate a faint outburst from SN\,2023ixf. In this analysis, we implicitly assume that any outburst would not have a timescale shorter than five days, which matches expectations for an optically thick eruption dominated by electron scattering opacity from a hydrogen-rich atmosphere \citep[$\kappa=0.2$~cm$^{2}$~g$^{-1}$, as in, e.g.,][]{Kleiser14}, a characteristic ejecta velocity of $v_{\rm ej}\approx$1000~km~s$^{-1}$ and ejecta masses $M_{\rm ej}>$0.001~$M_{\odot}$ such that the decline time is $t_{d} = \sqrt{\kappa M_{\rm ej} / (v_{\rm ej} c)}\gtrsim5$~days \citep[similar to analyses in][]{Dong2023, Ransome2024}.

We use the light curve shapes from \citet{Kuriyama20} for outbursts from RSGs in a similar mass range to SN\,2023ixf of 11~$M_{\odot}$ (i.e., their RSG1 model).  In general, the shape of these light curves exhibits a much longer timescale than five days, implying that our procedure would detect any emission with a well-defined peak luminosity.  We can scale these light curves to detect outbursts for an arbitrary ejecta mass, which is modeled as $\approx0.002$--0.5~$M_{\odot}$ of hydrogen-rich material in \citet{Kuriyama20}.  

\begin{figure}[bht!]
    \includegraphics[width=0.49\textwidth]{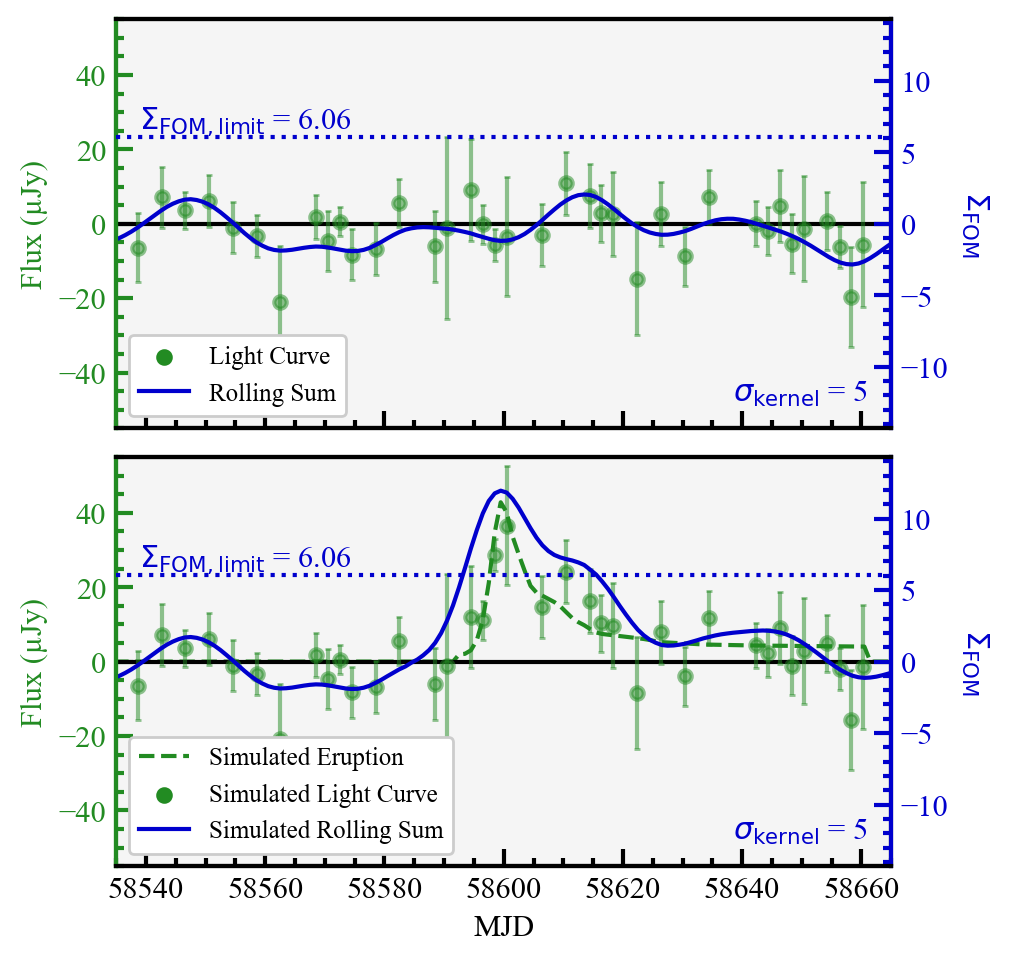}
    \caption{
        \textbf{Top}: In the green points, the example control light curve \#7 flux. The corresponding $\fom$ for $\sigmakern$ = 5 days in the solid blue line, and its $\fomlim$ = 6.06 in the dotted blue line. The $\fom$ does not cross $\fomlim$ at any point. \textbf{Bottom}: In the dashed green line, the shape of the simulated eruption peaking at $m_\text{peak}$ = 19.8 mag and 58600 MJD. In the green points, the resulting flux after injection of the simulated eruption. The corresponding simulated $\fom$ for $\sigmakern$ = 5 days in the solid blue line, and its $\fomlim$ = 6.06 in the dotted blue line. The $\fom$ crosses $\fomlim$ at approximately 58593 MJD.
    }
    \label{fig:simerup}
\end{figure}

Figure~\ref{fig:simerup} shows a simulated eruption with an ejecta mass of 0.05~$M_{\odot}$ and peaking at 19.8~mag, which we would expect to observe in our data.  Evidently, the control light curve $\fom$ successfully crosses the detection limit $\fomlim$ after injection of the simulated eruption.

\subsection{Efficiencies}

We determine the efficiency of our detection algorithm by calculating the detection success of a series of simulated eruptions within a certain parameter space. We analyze a list of different rolling Gaussian kernel sizes $\sigmakern$: 5, 15, 25, 40, 80, 130, 200, and 300 days. For each $\sigmakern$, we inject simulated Gaussian outbursts with similar kernel sizes $\sigmasim$, drawing the peak apparent magnitude $m_\text{peak}$ from a possible range of 16--23~mag and randomly drawing the peak MJD. Each simulated event is injected into a randomly drawn control light curve. We inject a total of 50000 simulated events for each $\sigmasim$. 

In order to determine whether a simulated event was detected, we scan the control light curve for $\fom > \fomlim$ located within $1\sigma$ of the peak MJD. Detections that fulfill these criteria are marked as successful. 

We can now calculate the efficiency of a certain $\sigmakern$ by determining how many of the simulated events with a certain $\sigmasim$ and $m_\text{peak}$ were successfully detected. For our main analysis we filter out any simulated events with a peak MJD located in any of the excluded MJD ranges described in Section~\ref{contameffmjdranges}. 

\begin{figure}[htb!]
\includegraphics[width=0.49\textwidth]{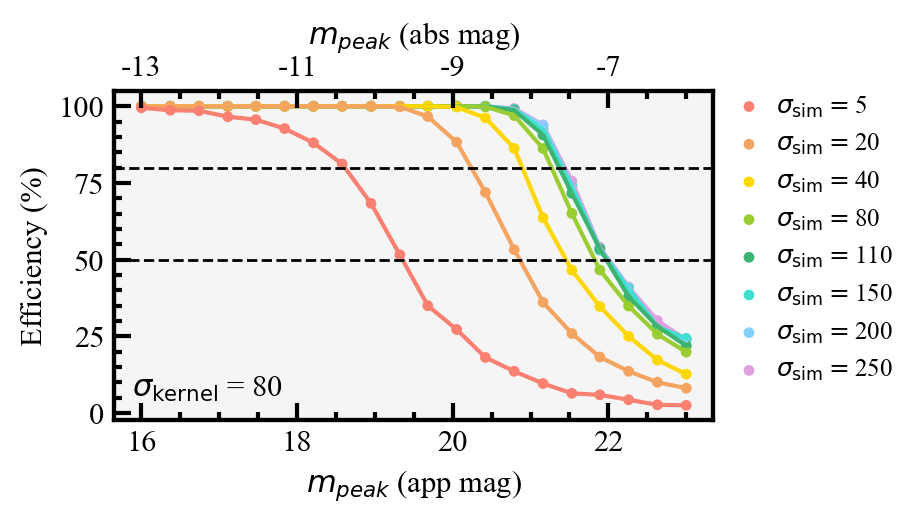}
    \caption{Efficiencies for $\sigmakern$ = 80 days and its detection limit $\fomlim$ = 23.63. We simulate events of kernel size $\sigmasim$ = 5, 20, 40, 80, 110, 150, 200, and 250 days and report the percentage of successful detections across peak apparent magnitudes from $m_{\text{peak}}$=23 to 16. The dashed lines at 50\% and 80\% efficiency demonstrate at which magnitude the efficiency curves cross these thresholds. }
    \label{fig:efficiency_80}
\end{figure}

After grouping the simulated events by $\sigmasim$ and peak apparent magnitude, we divide the number of successful detections by the total number of simulated events in that parameter space to get the efficiency. We display the efficiencies for an example kernel size of $\sigmakern$ = 80 days in Fig.~\ref{fig:efficiency_80}.

\begin{figure}[htb!]
\includegraphics[width=0.49\textwidth]{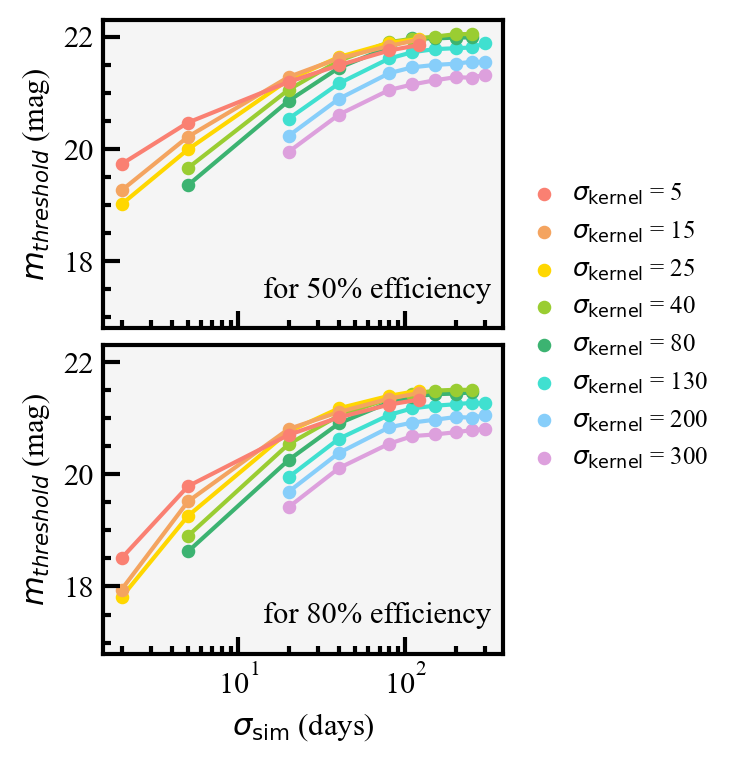}
    \caption{Apparent magnitude thresholds $m_{\text{threshold}}$ for 50\% (top panel) and 80\% efficiency (bottom panel) as a function of $\sigmasim$ on a log scale. Thresholds are displayed for each $\sigmakern$ using the corresponding detection limit $\fomlim$.}
    \label{fig:mag_thresholds}
\end{figure}

Finally, we obtain the apparent magnitude thresholds for 50\% efficiency and 80\% efficiency; that is, for each $\sigmakern$ and $\sigmasim$, we calculate the apparent magnitudes at which the efficiency curve crosses 50\% and 80\% (i.e., the magnitude thresholds at which we would detect transient emission 50\% and 80\% of the time, respectively). Figure~\ref{fig:mag_thresholds} illustrates all magnitude thresholds for each $\sigmakern$ across various corresponding $\sigmasim$. We list the deepest magnitude threshold and its corresponding $\sigmakern$ for each $\sigmasim$ in Table~\ref{table:best_mag_thresholds}. Appendix Section~\ref{mag_thresholds} further lists all magnitude thresholds as a function of $\sigmakern$ and $\sigmasim$.

\begin{table*}[htb!]
    \centering
    \caption{Best Magnitude Thresholds}
    \begin{tabular}{lcccc} 
\hline
    $\sigmasim$ & Best $\sigmakern$ & Best $m_\text{threshold}$ & Best $\sigmakern$ & Best $m_\text{threshold}$ \\
    \text{} & for 80\% efficiency & for 80\% efficiency & for 50\% efficiency & for 50\% efficiency \\
    \hline\hline
    2 & 5 & 18.50 & 5 & 19.73 \\
    5 & 5 & 19.79 & 5 & 20.47 \\
    20 & 15 & 20.80 & 5 & 21.19 \\
    40 & 25 & 21.17 & 5 & 21.49 \\
    80 & 25 & 21.40 & 5 & 21.75 \\
    110 & 40 & 21.43 & 40 & 21.96 \\
    150 & 40 & 21.49 & 40 & 22.00 \\
    200 & 40 & 21.50 & 40 & 22.04 \\
    250 & 40 & 21.50 & 40 & 22.04 \\
    300 & 130 & 21.26 & 130 & 21.88 \\
    \hline
    \end{tabular}
    \label{table:best_mag_thresholds}
\end{table*}

\section{Results \& Discussion}

Here we place our limits on pre-explosion optical emission from SN\,2023ixf in the context of potential progenitor models---specifically, detections of optical emission from a RSG undergoing variability and pre-explosion outbursts \citep{Kuriyama20,Davies22} or observed from SN\,2020tlf \citet{jacobson-galan2022}.  We compare our results to previous studies of SN\,2023ixf in \citet{Dong2023} as well as direct detections of the progenitor star from the optical to mid-IR \citep[e.g.,][]{jencson2023,kilpatrick2023,soraisam2023,vandyk2023}.

\subsection{Constraints on Optical Variability from the SN~2023ixf Progenitor}

The RSG progenitor stars to Type\,II SNe are primarily detected in F814W (roughly $I$-band, with effective wavelength 7960~\AA), which in general is a redder band and therefore brighter than RSGs with similar spectral types compared with ATLAS $o$-band.  \citet{davies2018} compiled F814W detections of SN\,II progenitor star detections with unextinguished F814W absolute magnitudes from $-6.1$ to $-8.4$~mag, or 20.8 to 23.1~mag at the distance and extinction to M101.  For comparison, SN\,2023ixf exhibited apparent magnitudes of $m_{\rm F814W}=24.9$~mag and $m_{\rm F675W}=26.4$~mag at approximately 24 to 19 years prior to explosion \citep[see][]{Kilpatrick2023a}, consistent with expectations for a highly reddened RSG surrounded by CSM \citep{jencson2023,kilpatrick2023,soraisam2023}.  If this RSG went through periods of extreme optical activity or weak mass loss such that its optical-IR colors were low, it may have been detectable in ATLAS $o$-band in our 2015 to 2023 observations (see Table~\ref{table:best_mag_thresholds} for our magnitude thresholds).

We test this hypothesis below by injecting sources with constant flux into our ATLAS $o$-band data from 2015--2023 for a rolling sum with a width of $\sigmakern=300$~days.  This width corresponds to approximately 1/4 the variability timescale of the SN\,2023ixf as constrained by near- and mid-IR observations and so reflects a scenario where the optical counterpart is brighter than our limit for approximately half of this timescale.  Using this timescale, we rule out sources with $m_{o}<21.3$~mag with 80\% efficiency, which we adopt as our fiducial limit on the average brightness of a source with long timescale variability below.  We note here that because we estimate this limit in difference imaging, it corresponds to a limit on the differential flux of any source that could appear in the data.  Below, we implicitly assume that there is effectively zero source flux in the template image, which allows us to treat our magnitude limit as a constraint on the stellar counterpart to SN\,2023ixf.

\begin{figure}[htb]
    \includegraphics[width=0.48\textwidth]{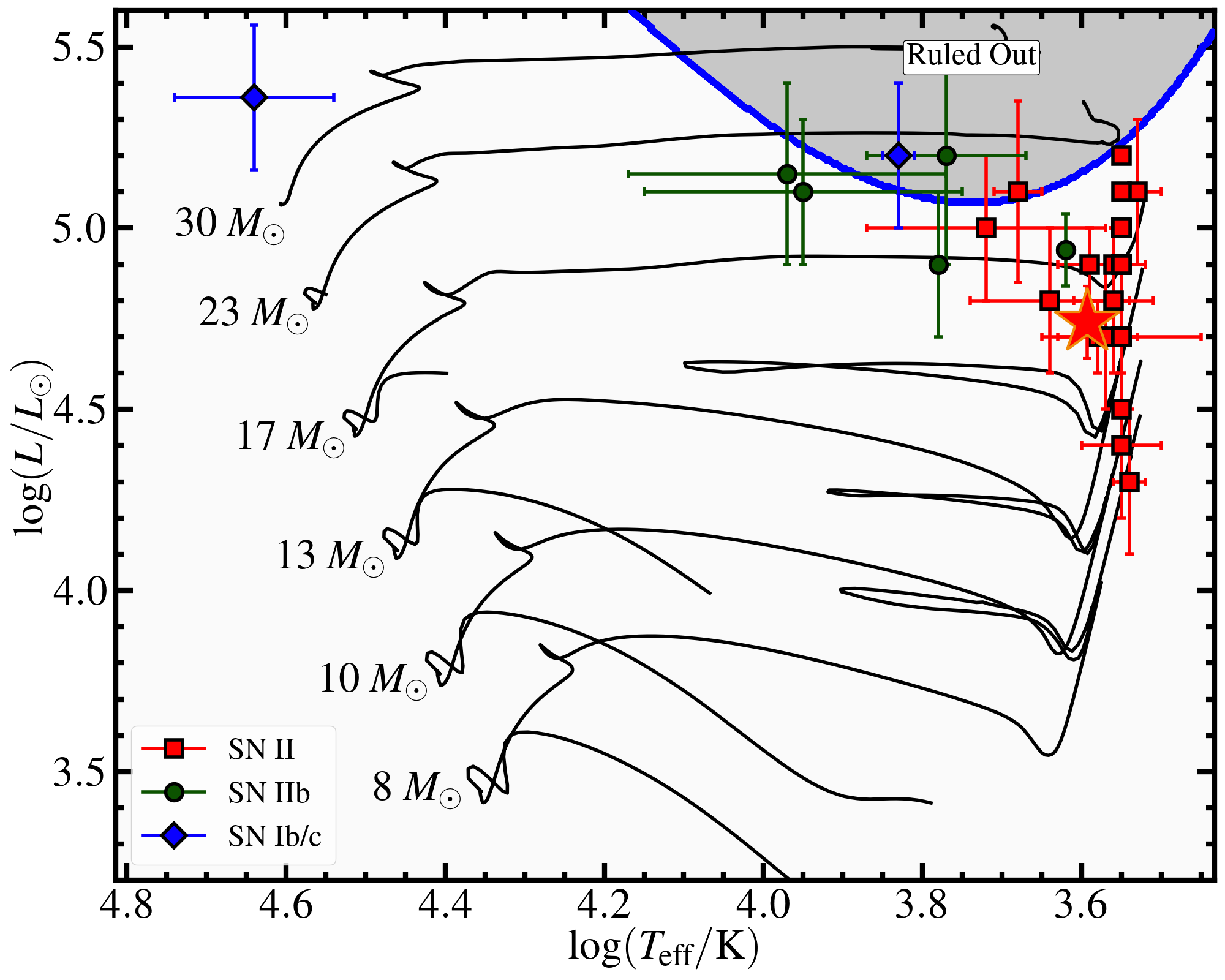}
    \caption{Hertzsprung-Russell diagram showing the region of parameter space ruled out by our ATLAS $o$-band limits for $\sigmakern=300$~day and assuming a blackbody spectrum.  For comparison, we show Mesa Isochrone \& Stellar Evolution Tracks \citep[MIST;][]{choi16} models for single stars with initial masses from 8--30~$M_{\odot}$ as well as the locations of SN\,II, IIb, and Ib progenitor stars \citep[from][and references therein]{smartt2015,kilpatrick2021}.  The location of the SN\,2023ixf progenitor star (without circumstellar extinction) from \citet{kilpatrick2023} is shown as a red star.  Our ATLAS $o$-band limit (21.3~mag) rules out massive and luminous RSGs at $>$22~$M_{\odot}$ but does not extend to less massive systems.}\label{fig:hr}
\end{figure}

This limit roughly corresponds to the terminal state of the most luminous RSGs in Mesa Isochrone \& Stellar Tracks \citep[MIST;][]{choi16} models with $M_{\rm ZAMS}\approx22~M_{\odot}$ and without any additional circumstellar extinction (Fig.~\ref{fig:hr}).

The limit calculated above is comparable to the largest reported masses for the SN\,2023ixf counterpart \citep[ranging from 9--21~$M_{\odot}$ depending on assumptions of rotation and bolometric corrections;][]{jencson2023,kilpatrick2023,soraisam2023,vandyk2023}, although SN\,2023ixf exhibited evidence for extreme optical extinction ($A_{V}>4$~mag) in modeling of its pre-explosion counterpart.  It is theoretically plausible that if the counterpart went through a period of weak mass loss and its optical emission was significantly enhanced \citep[e.g., similar to optical emission from $\alpha$ Orionis with a similar initial mass;][]{Taniguchi22} that our ATLAS data would be sensitive to this emission.  We therefore rule out the most extreme systems where the counterpart was both extremely luminous and underwent evolution with extremely low circumstellar extinction.

In the future, ATClean can be used to search for variability over arbitrary timescales in imaging from the Vera C. Rubin Observatory.  In particular, the single epoch $z$ and $y$-band limits for extragalactic star clusters are expected to be 22.79 and 22.00~mag, respectively \citep{Usher2023}.  Rubin will be effective at constraining variability from extragalactic RSGs in these bands, and application of ATClean will allow us to distinguish between true variability and anomalous detections down to the expected limit of its imaging.  Even unextinguished RSGs are $\approx$1.7~mag brighter in $z$-band and 2.1~mag brighter in $y$-band relative to $o$-band.  Combined with the deep limits from Rubin, we expect to be able to detect RSG variability down to the least massive $\approx$7.5~$M_{\odot}$ initial mass RSG stars.

\subsection{Constraints on Pre-explosion Outbursts from SN~2023ixf}

In addition to long-period variability from massive stars, pre-explosion outbursts are predicted for SN\,II progenitor systems where rapid energy injection into the progenitor star's envelope \citep[e.g., a ``nuclear flash'' or other burning instabilities;][]{Woosley78,Dessart10} can lead to a non-terminal eruption.  This mechanism is ordinarily invoked for precursors to Type IIn SNe \citep[SN\,2000ch, 2009ip, 2010mc, 2016blu, 2016jbu;][]{Pastorello10,Smith11,Kilpatrick18,Aghakhanloo23a, Aghakhanloo23b}, but the observation of the otherwise ``normal'' Type II-P SN\,2020tlf suggests this mechanism extends to SNe with lower mass RSG progenitor systems similar to SN\,2023ixf \citep{jacobson-galan2022}.


The high temporal sampling of our ATLAS limits combined with our novel method for investigating precursor emission enables us to place unique constraints on pre-explosion emission from SN\,2023ixf \citep[similar to][]{Dong2023}.  Setting $\sigmasim$ to a relatively short timescale with $\sigmasim\approx5$~days, we consider the prescription of \citet{Kuriyama20} in transforming our limiting magnitudes into a physically meaningful constraint on ejected mass.  For this timescale and set of models, we find that the maximum brightness of a precursor outburst from SN\,2023ixf would be $o=19.8$~mag for a recovery efficiency of 80\%.  This is comparable to the limit of $o=20.0$~mag ($M_{o}=-9.2$~mag) reported in \citet{Dong2023}, but we also use a moderately shorter convolution timescale (five as opposed to 10~days) to better match the expected duration of light curves from \citet{Kuriyama20}. 

Taking into consideration that the ATLAS cadence is 1-2 days \citep{tonry2018}, however, events with timescales close to this cadence cannot be sampled well and are intrinsically more difficult to detect than those with longer timescales. We face a similar dilemma with events with larger timescales approaching $\sigmasim$ = 300 days, as ATLAS observation seasons tend to be approximately 270 days. As shown in Table~\ref{table:best_mag_thresholds}, $\sigmasim$ of 300 days reaches a threshold of 21.26 for $\sigmakern$ = 130 days, and $\sigmasim$ from 80 to 250 days reach approximately 21.4-5 mag for $\sigmakern$ = 40 days. We can conclude that events with timescales from 80 to 300 days reach 21.3 mag or deeper for a recovery efficiency of 80\%, with varying corresponding best $\sigmakern$ per Table~\ref{table:best_mag_thresholds}.



Based on the adopted model, we can rule out mass ejections of $>$0.021~$M_{\odot}$ for nearly the entire duration of our ATLAS observations, much lower than the confined CSM mass of 0.04--0.07~$M_{\odot}$ observed around SN\,2023ixf from early optical photometry and spectroscopy \citep[][]{bostroem23,hiramatsu2023,jacobson-galan2023,smith2023}.  This suggests that an eruptive mass-loss mechanism is unlikely as the source of this material, and instead a steady but extremely enhanced mass-loss rate \citep[i.e., over normal mass-loss prescriptions for RSGs such as in][]{Beasor20} is a more likely explanation \citep[e.g., a ``superwind'' as in][]{Davies22}.  However, our method can be used to systematically search for such precursor outbursts in current and future optical surveys with deep, sufficiently high-cadence imaging.

\section{Conclusions}

In order to reliably determine the presence of pre-SN eruptions in SN\,2023ixf, we  processed and obtained high-fidelity binned ATLAS light curves.  We placed deep limits on the presence of pre-SN outbursts using a novel method for quantifying the presence of faint signals in forced photometry light curves from time-domain surveys called ATClean.  In summary, we find:

\begin{enumerate}
    \item ATClean reliably reproduces 3$\sigma$ flux limits in well-sampled, high-cadence ATLAS difference imaging forced photometry while avoiding anomalous detections of transient emission due to instrument artifacts, contamination from bright sources of background emission, or artifacts introduced from the image reduction and subtraction procedure.
    \item We do not detect any significant sources of transient emission at the site of SN\,2023ixf over timescales from 5--300~days in pre-explosion ATLAS $o$-band data.  Our nominal magnitude limits where we would detect transient emission 80\% of the time (i.e., with 80\% efficiency) are 19.8~mag over 5~days and 21.3~mag over 80 to 300~days.
    \item Based on the five day timescale limit, we rule out outbursts that would yield $>$0.021~$M_{\odot}$ of ejecta, comparable to limits on outbursts in \citet{Dong2023}.  The latter limits rule out the most luminous $>$22~$M_{\odot}$ RSGs with no additional circumstellar extinction based on comparison to MIST single-star evolution models.  In the future this procedure can be applied to search for variability in Rubin $z$ and $y$-band imaging to search for variability from even the least massive RSGs.
\end{enumerate}

\section*{Acknowledgements}
We are very thankful for the support of AR, JJ, and XL by NSF AST grants 1814993 and 2108841. DC and QW are partially supported by NASA ADAP grant 80NSSC22K0494 and the STScI DDRF fund. This work was funded by ANID through grant ICN12\_12009 awarded to the Millennium Institute of Astrophysics (MAS).

\pagebreak
\appendix

\section{Bisection Method for Detection Limit Determination}
\label{bisectionmethod}

\begin{algorithm}[H]
\SetKwInOut{Input}{input}\SetKwInOut{Output}{output}
\Input{contamination function $g$, maximum iterations $N_{\text{max}}$, lower endpoint value $\beta$, upper endpoint value $\alpha$}
\Output{value $\gamma$, $\alpha \geq \gamma \geq \beta$, with target contamination value $g(\gamma) = 1$}
 $N \leftarrow 0$\;
 
\While{$N < N_{\text{max}}$}{
    $\gamma \leftarrow \frac{\alpha + \beta}{2}$\;
    \eIf{$g(\gamma) > 1$}{
        $\beta \leftarrow \gamma$\;
    }{
        $\alpha \leftarrow \gamma$\;
    }    
    $N \leftarrow N + 1$\;
}
\caption{Bisection Method}
\end{algorithm}

\section{Alternative Method for Detection Limit Determination}
\label{deteclimitalt}

\begin{figure}[htb!]
    \centering
    \includegraphics[width=0.7\linewidth]{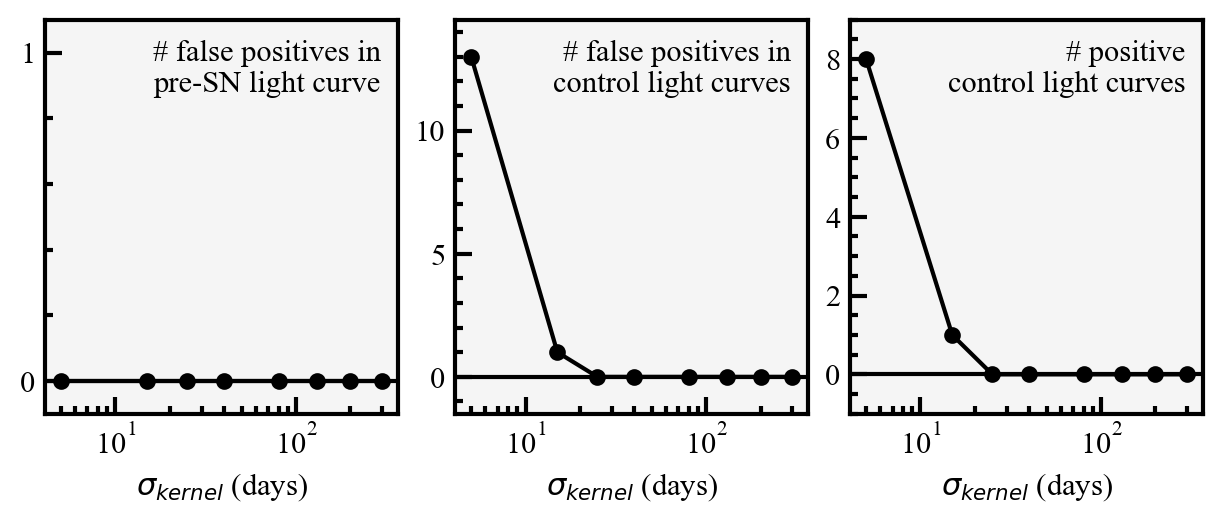}
    \caption{Contamination for each $\sigmakern$ in terms of the total number of false positives in the pre-SN light curve (left panel), the total number of false positives in the control light curves (middle panel), and the number of control light curves with false positives (right panel). Contamination is calculated using the detection limit $\fomlim$ obtained by scaling the standard deviation of the $\fom$ distribution.}
    \label{fig:contam_3sigma}
\end{figure}

As described in Section~\ref{deteclimits}, we calculated a preliminary detection limit $\fomlim$ for each rolling Gaussian kernel size $\sigmakern$ by scaling the standard deviation of the corresponding $3\sigma$ clipped $\fom$ distribution. However, due to the inconsistent shape of the $\fom$ distributions for larger kernel sizes, these detection limits become decreasingly meaningful as the kernel size increases. To paint a better picture of the (lack of) efficacy of these detection limits, we measure contamination in three different ways: a) the total number of false positives in the pre-SN light curve, b) the total number of false positives in the control light curves, and c) the number of control light curves with any false positives. In Fig.~\ref{fig:contam_3sigma}, the smaller kernel sizes clearly exhibit more contamination and the larger kernel sizes exhibit less contamination across two of the three measures. 

\begin{figure}[htb!]
    \centering
    \includegraphics[width=0.5\linewidth]{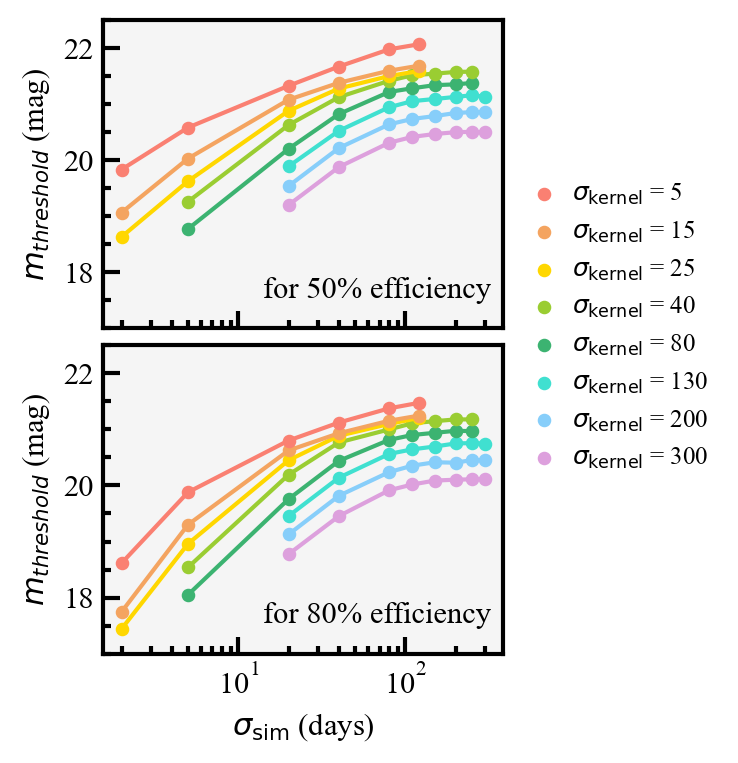}
    \caption{Apparent magnitude thresholds $m_{\text{threshold}}$ for 50\% (top panel) and 80\% efficiency (bottom panel) as a function of $\sigmasim$ on a log scale. Thresholds are displayed for each $\sigmakern$ using the detection limit $\fomlim$ obtained by scaling the standard deviation of the $\fom$ distribution.}
    \label{fig:mag_thresholds_3sigma}
\end{figure}

Crucially, the efficiency of a kernel size depends heavily on the strictness of its detection limit. Therefore, the previously demonstrated asymmetry in contamination implies that the efficiencies of larger kernel sizes could be negatively impacted, and those of smaller kernel sizes inflated. Fig.~\ref{fig:mag_thresholds_3sigma} displays the apparent magnitude thresholds for 50\% and 80\% efficiency using the discussed detection limits. The smaller kernel sizes consistently exhibit deeper magnitude thresholds than the larger kernel sizes, even for larger $\sigmasim$. 

\section{Excluding MJD Ranges}
\label{s2}

In Table~\ref{table:mjdranges}, the start MJD and end MJD for each excluded MJD range described in Section~\ref{contameffmjdranges} is provided. These excluded ranges consist of the first two observation seasons and the gaps between observation seasons. We additionally pad these gaps with extra excluded days if they exhibit high numbers of non-detected bright (peak apparent magnitude $m_\text{peak} \leq$18 mag) simulations, as we aim for an equal distribution of failed detections over time to obtain optimal efficiencies. Furthermore, we determine a magnitude threshold for each efficiency curve in our main analysis by calculating the peak apparent magnitudes $m_\text{peak}$ at which the efficiency curve first reaches 50\% and 80\%; therefore, bright simulated events must reach approximately 100\% efficiency in order for our magnitude thresholds to be an accurate representation of the search depth of our detection algorithm.

\begin{figure*}[htb!]
    \centering
    \includegraphics[width=\linewidth]{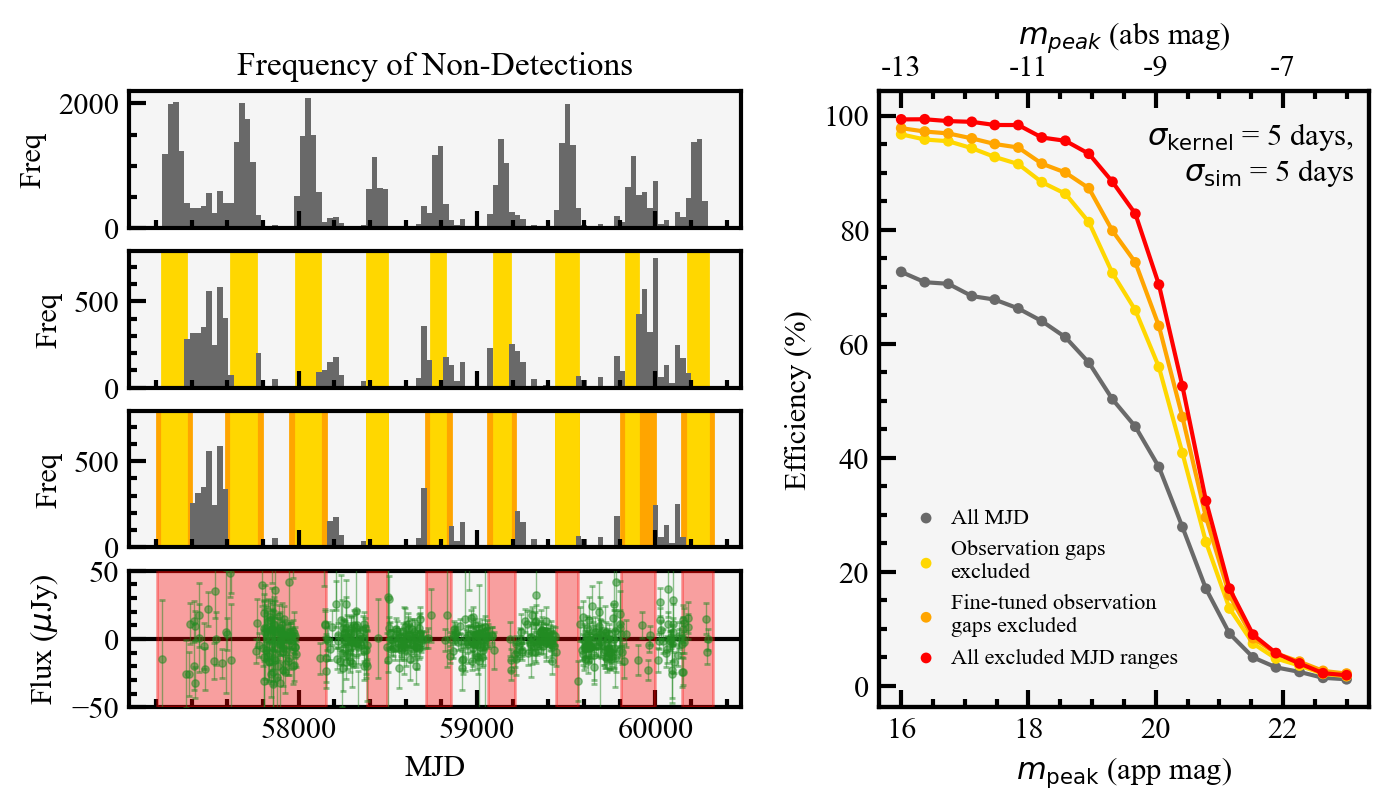}
    \caption{\textbf{Left}: In the top panel, the distribution of peak MJDs for non-detected simulated events with $m_\text{peak} \leq$18 mag. In the second panel down, the distribution of any remaining non-detections occurring outside of the observation gaps (yellow). In the third panel down, the distribution of any remaining non-detections occurring outside of the fine-tuned observation gaps (yellow and orange). In the bottom panel, the example control light curve \#7 with all excluded MJD ranges (the fine-tuned observation gaps \textit{and} the problematic first two observation seasons) highlighted in red. \textbf{Right}: The efficiencies for $\sigmakern$ = 5 days and $\sigmasim$ = 5 days illustrated for different MJD ranges. In gray, the efficiency for all simulated events. In yellow, the efficiency for simulated events with peak MJDs outside of the observation gaps. In orange, the efficiency for simulated events with peak MJDs outside of the fine-tuned observation gaps. In red, the efficiency for simulated events with peak MJDs outside of all excluded MJD ranges (the fine-tuned observation gaps \textit{and} the problematic first two observation seasons).}
    \label{fig:mjd_ranges_hist2_5}
\end{figure*}

In Fig.~\ref{fig:mjd_ranges_hist2_5}, we depict the distribution of non-detections over time, as well as how our excluded MJD ranges effectively cover high-frequency areas of non-detections. We additionally show example efficiency curves for simulations across all MJD, outside the observation gaps, outside the fine-tuned observation gaps, and outside all excluded MJD ranges (the fine-tuned observation gaps \textit{and} the problematic first two observation seasons). Evidently, our complete set of all excluded MJD ranges results in approximately 100\% efficiency for bright simulations.


\begin{table*}[htb!]
    \centering
    \caption{Excluded MJD Ranges}
    \begin{tabular}{cc} 
\hline
Start MJD & End MJD \\
    \hline\hline
    57203 & 58150 \\
    58383 & 58494 \\
    58711 & 58852 \\
    59063 & 59214 \\
    59445 & 59566 \\
    59805 & 60001 \\
    60152 & 60325 \\
    \hline
    \end{tabular}
    \label{table:mjdranges}
\end{table*}

\section{Apparent Magnitude Thresholds}
\label{mag_thresholds}

\smallskip

\startlongtable
\begin{deluxetable*}{ccccc}
\tablecaption{Magnitude Thresholds for Gaussian Light Curves Analyzed with Given $\sigmakern$ and $\sigmasim$}
\tablehead{
$\sigmakern$ &
$\sigmasim$ &
$\fomlim$ &
$m_\text{threshold}$ for 50\% efficiency &
$m_\text{threshold}$ for 80\% efficiency \\
\textbf{(days)} &
\textbf{(days)} &
&
\textbf{(mag)} &
\textbf{(mag)}
}
\startdata
5 & 2 & 6.06 & 19.73 & 18.50 \\
5 & 5 & 6.06 & 20.47 & 19.79 \\
5 & 20 & 6.06 & 21.19 & 20.70 \\
5 & 40 & 6.06 & 21.49 & 21.01 \\
5 & 80 & 6.06 & 21.75 & 21.24 \\
5 & 120 & 6.06 & 21.85 & 21.33 \\
15 & 2 & 9.68 & 19.26 & 17.94 \\
15 & 5 & 9.68 & 20.21 & 19.51 \\
15 & 20 & 9.68 & 21.28 & 20.80 \\
15 & 40 & 9.68 & 21.61 & 21.11 \\
15 & 80 & 9.68 & 21.83 & 21.35 \\
15 & 120 & 9.68 & 21.94 & 21.44 \\
25 & 2 & 12.27 & 19.01 & 17.81 \\
25 & 5 & 12.27 & 19.99 & 19.26 \\
25 & 20 & 12.27 & 21.23 & 20.76 \\
25 & 40 & 12.27 & 21.64 & 21.17 \\
25 & 80 & 12.27 & 21.89 & 21.40 \\
25 & 120 & 12.27 & 21.95 & 21.49 \\
40 & 5 & 16.27 & 19.66 & 18.90 \\
40 & 20 & 16.27 & 21.05 & 20.54 \\
40 & 40 & 16.27 & 21.57 & 21.04 \\
40 & 80 & 16.27 & 21.90 & 21.35 \\
40 & 110 & 16.27 & 21.96 & 21.43 \\
40 & 150 & 16.27 & 22.00 & 21.49 \\
40 & 200 & 16.27 & 22.04 & 21.50 \\
40 & 250 & 16.27 & 22.04 & 21.50 \\
80 & 5 & 21.76 & 19.35 & 18.62 \\
80 & 20 & 21.76 & 20.85 & 20.25 \\
80 & 40 & 21.76 & 21.44 & 20.90 \\
80 & 80 & 21.76 & 21.81 & 21.28 \\
80 & 110 & 21.76 & 21.97 & 21.38 \\
80 & 150 & 21.76 & 21.96 & 21.43 \\
80 & 200 & 21.76 & 21.98 & 21.44 \\
80 & 250 & 21.76 & 21.98 & 21.46 \\
130 & 20 & 28.31 & 20.53 & 19.95 \\
130 & 40 & 28.31 & 21.16 & 20.63 \\
130 & 80 & 28.31 & 21.61 & 21.06 \\
130 & 110 & 28.31 & 21.72 & 21.17 \\
130 & 150 & 28.31 & 21.78 & 21.22 \\
130 & 200 & 28.31 & 21.79 & 21.25 \\
130 & 250 & 28.31 & 21.80 & 21.27 \\
130 & 300 & 28.31 & 21.88 & 21.26 \\
200 & 20 & 36.81 & 20.23 & 19.69 \\
200 & 40 & 36.81 & 20.89 & 20.38 \\
200 & 80 & 36.81 & 21.35 & 20.83 \\
200 & 110 & 36.81 & 21.45 & 20.92 \\
200 & 150 & 36.81 & 21.49 & 20.97 \\
200 & 200 & 36.81 & 21.52 & 21.02 \\
200 & 250 & 36.81 & 21.55 & 21.01 \\
200 & 300 & 36.81 & 21.54 & 21.05 \\
300 & 20 & 47.14 & 19.95 & 19.41 \\
300 & 40 & 47.14 & 20.61 & 20.11 \\
300 & 80 & 47.14 & 21.05 & 20.54 \\
300 & 110 & 47.14 & 21.15 & 20.68 \\
300 & 150 & 47.14 & 21.22 & 20.71 \\
300 & 200 & 47.14 & 21.28 & 20.75 \\
300 & 250 & 47.14 & 21.26 & 20.78 \\
300 & 300 & 47.14 & 21.32 & 20.80 \\
\enddata
\end{deluxetable*}

\section{Supplementary Figures}

\begin{figure}[p]
    \centering
    \includegraphics[scale=0.85]{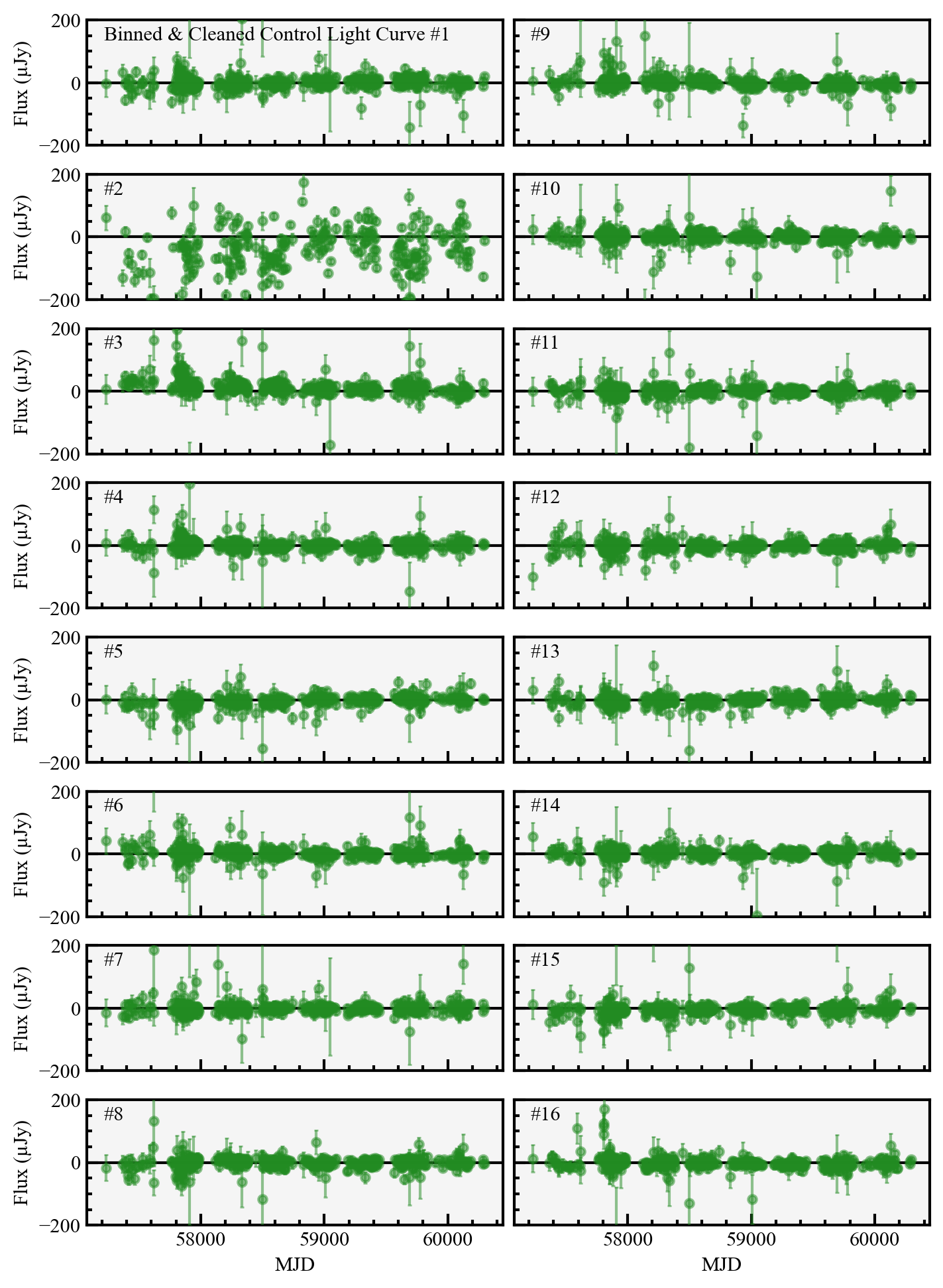}
    \caption{Binned and cleaned ATLAS $o$-band light curves for SN~2023ixf's 16 control positions, labeled by ID \#1-16.}
    \label{fig:all_controls}
\end{figure}

\begin{figure}[p]
    \centering
    \includegraphics[scale=0.62]{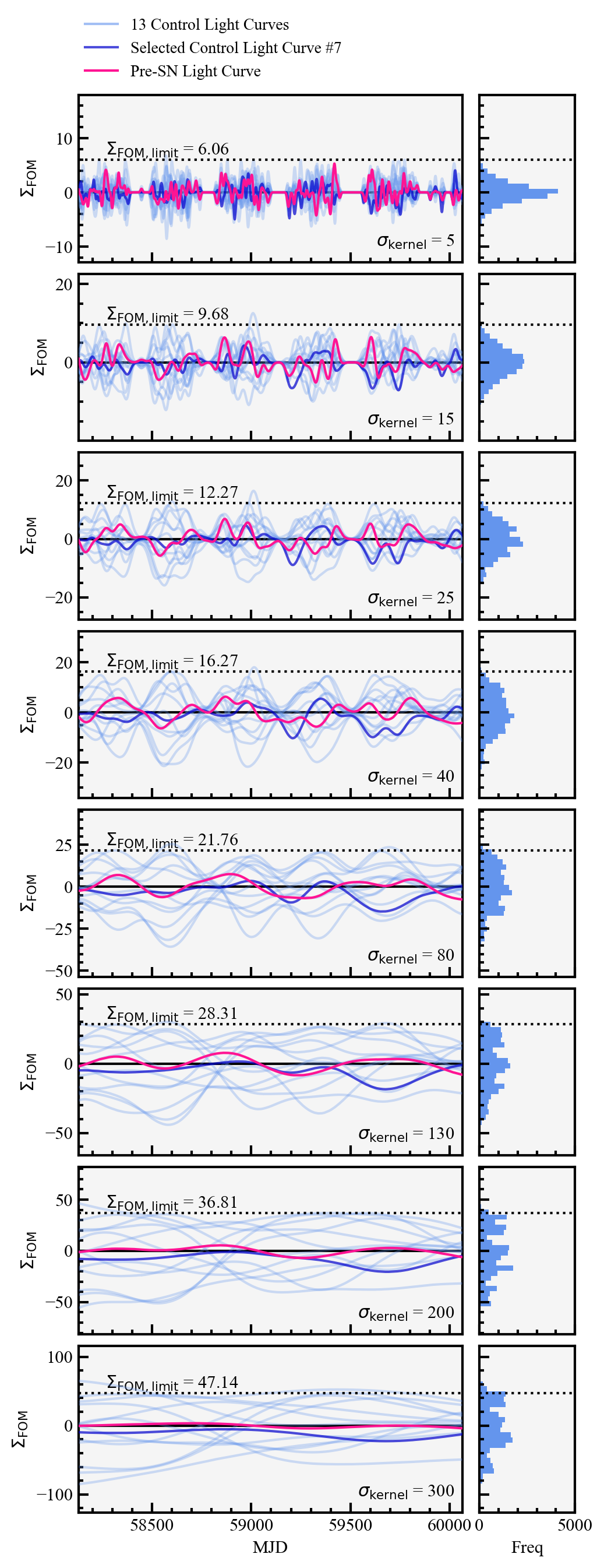}
    \caption{From top to bottom, the $\fom$ of the light curves with $\sigmakern$ = 5, 15, 25, 40, 80, 130, 200, and 300 days, and the corresponding $\fomlim$ in the dotted black lines. To the right, the control light curve $\fom$ distributions for the corresponding $\sigmakern$.}
    \label{fig:all_foms}
\end{figure}

Fig.~\ref{fig:all_controls} depicts all sixteen binned and cleaned control light curves. Fig.~\ref{fig:all_foms} depicts the control light curve and pre-SN light curve $\fom$ for the complete set of $\sigmakern$ used in our main analysis.

\begin{figure}[p]
\includegraphics[width=\textwidth]{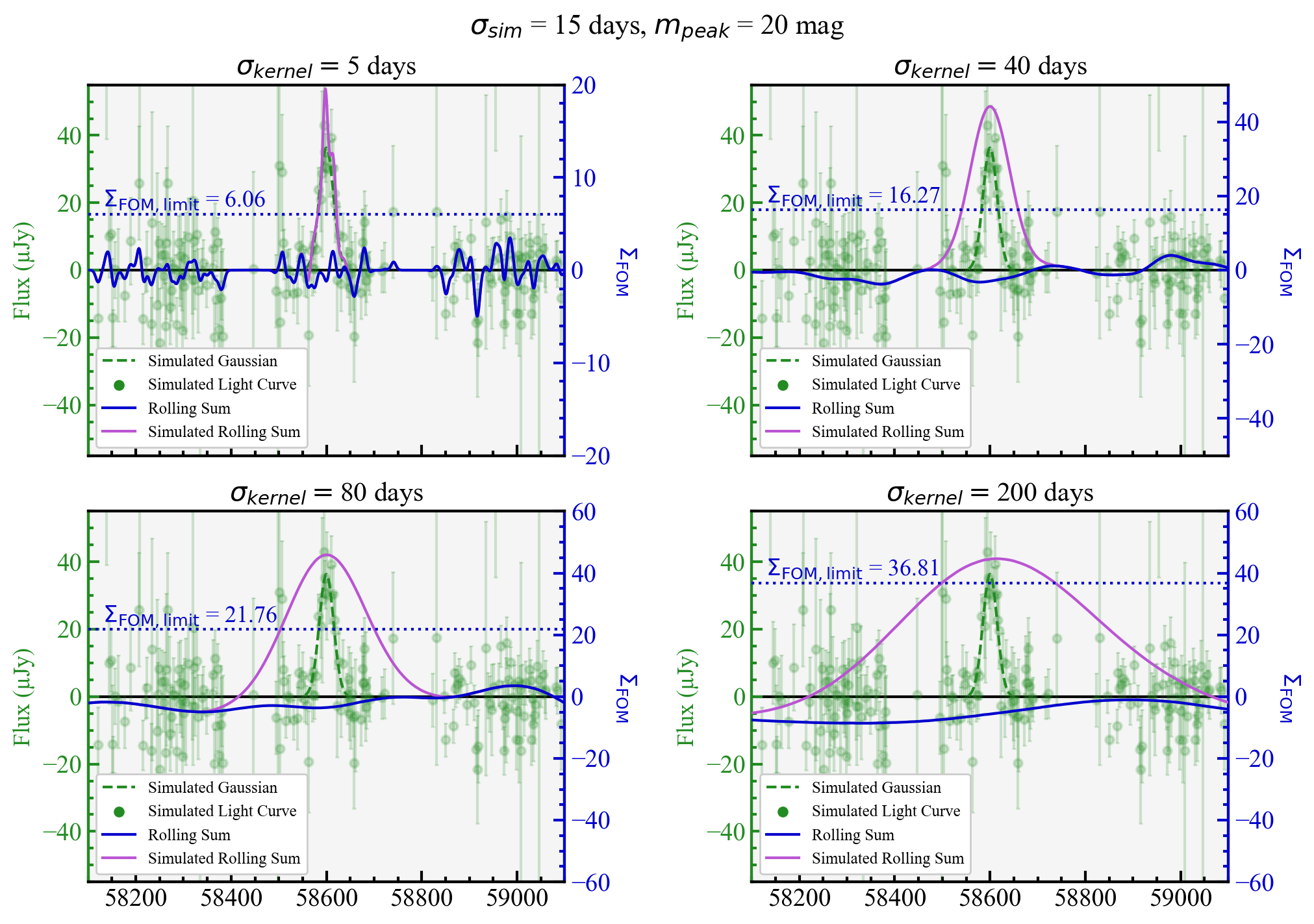}
    \caption{The same short-timescale simulated event convolved with rolling Gaussians of different kernel sizes: $\sigmakern$ = 5 days (top left), 40 days (top right), 80 days (bottom left), and 200 days (bottom right). In the dashed green line, the shape of the simulated Gaussian with $\sigmasim$ = 15 days and $m_\text{peak}$ = 20 mag. In the green points, the flux of example control light curve \#7 after injection of the simulated event. The original $\fom$ in the solid blue line, and the $\fom$ convolved with the simulated flux in the solid purple line. The simulated $\fom$ successfully crosses the corresponding $\fomlim$ (dotted blue line) for each rolling Gaussian kernel size $\sigmakern$.}
    \label{fig:simbumps_15}
\end{figure}

\begin{figure}[p]
\includegraphics[width=\textwidth]{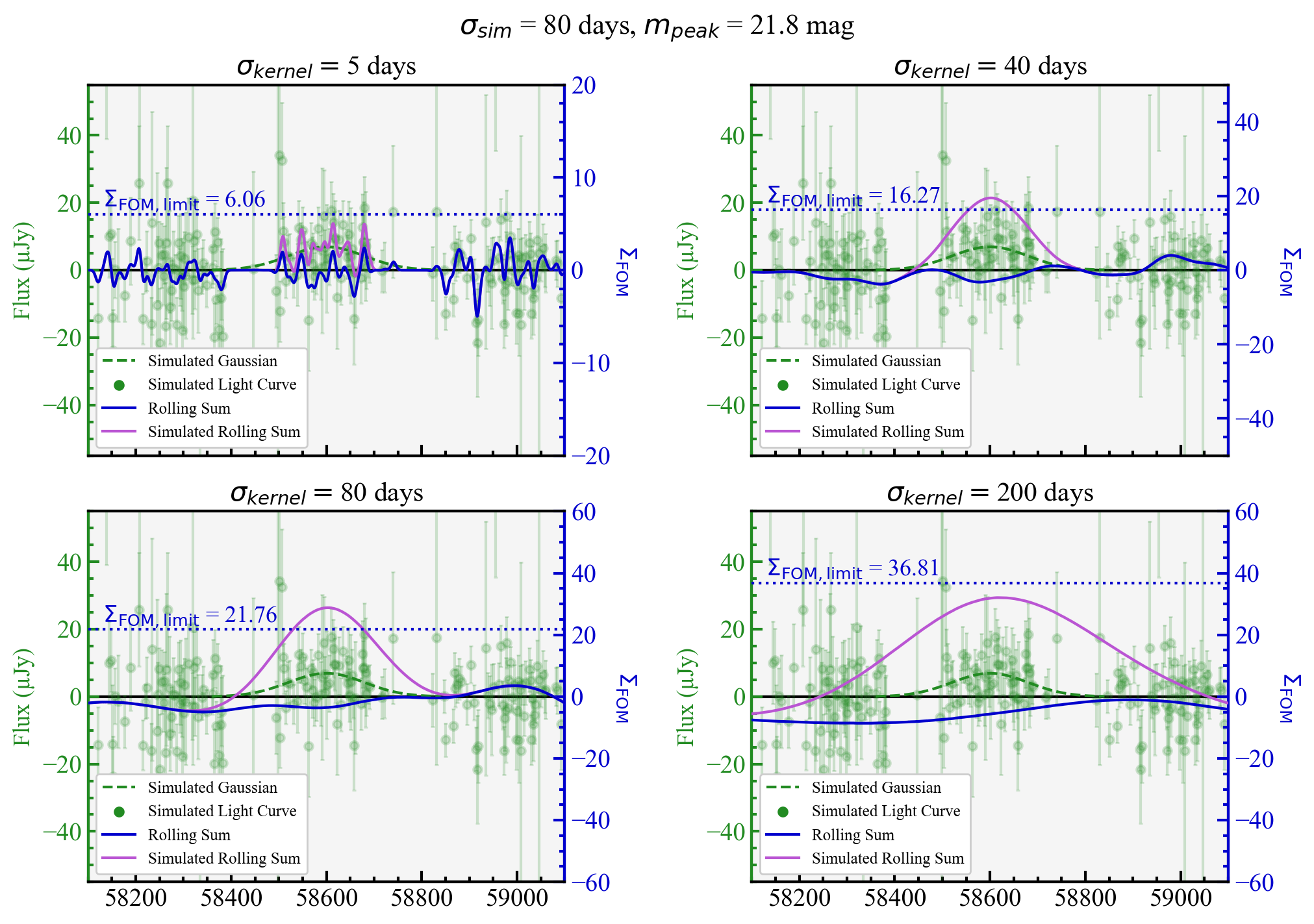}
    \caption{The same faint long-timescale simulated event convolved with rolling Gaussians of different kernel sizes: $\sigmakern$ = 5 days (top left), 40 days (top right), 80 days (bottom left), and 200 days (bottom right). In the dashed green line, the shape of the simulated Gaussian with $\sigmasim$ = 80 days and $m_\text{peak}$ = 21.8 mag. In the green points, the flux of example control light curve \#7 after injection of the simulated event. The original $\fom$ in the solid blue line, and the $\fom$ convolved with the simulated flux in the solid purple line. The simulated $\fom$ successfully crosses the corresponding $\fomlim$ (dotted blue line) for $\sigmakern$ = 40 and 80 days.}
    \label{fig:simbumps_80}
\end{figure}

In Fig~\ref{fig:simbumps_15}, we inject the same short-timescale simulated event with $\sigmasim$ = 15 days and $m_{\text{peak}}$ = 20 mag to the same location in the example control light curve, then convolve the resulting flux with four rolling Gaussians of different kernel sizes. The simulated event crosses the detection limit for each kernel size, although to a greater degree for smaller kernel sizes than for larger kernel sizes. Similarly, we inject a faint longer-timescale simulated event with $\sigmasim$ = 80 days and $m_{\text{peak}}$ = 21.8 mag and convolve it in Fig.~\ref{fig:simbumps_80}. The simulated event only crosses the detection limit when convolved with kernel sizes $\sigmakern$ = 40 and 80 days.

\bibliography{references}{}
\bibliographystyle{aasjournal}

\end{document}